\documentclass[aps,pre,epsf,showpacs,showkeys,epsf,floats,preprint]{revtex4}
\usepackage{amsmath}
\usepackage{amsfonts}
\usepackage[dvipdfm]{graphics}
\usepackage{bm}
\begin{document}
\title{Control of Integrable Hamiltonian Systems and Degenerate Bifurcations}
\author{C.W. Kulp \& E.R. Tracy}
\affiliation{Physics Department, College of William \& Mary, Williamsburg, VA
23187-8795}
\date{}
\begin{abstract}
We discuss control of low-dimensional systems which, when uncontrolled, are integrable in the Hamiltonian sense.  The controller targets an exact solution of the system in a region where the uncontrolled dynamics has invariant tori.  Both dissipative and conservative controllers are considered. We show that the shear flow structure of the undriven system causes a Takens-Bogdanov birfurcation to occur when control is applied.  This implies extreme noise sensitivity.  We then consider an example of these results using the driven nonlinear Schr\"{o}dinger equation.
\end{abstract}
\maketitle

\section{Introduction}\label{intro}
In the last several years there has been a lot of interest in improving the methods for the control of nonlinear physical systems.  Due to the large variety of behaviors which nonlinear systems display, a single control theory which can be applied to all nonlinear systems will be very difficult (if not impossible) to develop.  It seems likely that the best way to approach the control of nonlinear systems is to learn how to control classes of systems instead of trying to develop one single all-encompassing theory.  For example, in~\cite{ortega} control of Euler-Lagrange systems, which encompass a very large class of nonlinear physical systems, are considered.  Similarly, the work reported here restricts itself to a particular class of physical systems:  those which are well modeled by integrable, or near-integrable, Hamiltonian dynamical models.  Finite dimensional examples of integrable systems are linear oscillators and certain systems of nonlinear oscillators.  Examples in infinite dimensions includes soliton systems such as the Korteweg-de Vries (KdV), nonlinear Schr\"{o}dinger (NLS), and sine-Gordon equations.  In particular we are interested in studying the interplay of dissipative and conservative terms as a means to control integrable Hamiltonian systems.  Toward this end, our strategy is to use a known exact solution as a target by turning it into an attractor, which cannnot exist in the original Hamiltonian dynamics.  It is hoped that the knowledge gained from developing control laws for integrable Hamiltonian models that are simple to use, and robust to perturbations, will provide insights for developing control laws for real physical systems.  The main result of this paper is that when a seemingly natural controller is applied to integrable Hamiltonian systems, a highly degenerate bifurcation known as a Takens-Bogdanov bifurcation occurs which severely limits the controllability of the system. As an example, the results described in Sections~\ref{intro}-~\ref{takens} are then illustated on the nonlinear Schr\"{o}dinger (NLS) equation.  The NLS is integrable in the Hamiltonian sense and is a model system used to study phenomenon in plasmas and nonlinear optics and a variety of other fields.  

Here it is worth mentioning that Vaidya and Mezi\'{c}~\cite{umesh} have studied the controllability of a class of area-preserving twist maps.  These twist maps are one-dimensional integrable Hamiltonian systems.  They show that, under certain conditions, when a \textit{time-dependent} controller is applied to the integrable twist-map global controllability can be attained for the system.  In other words, the system can be controlled from any initial state to any final state.   Although their work deals with maps and not flows, this demonstrates that \textit{global} controllability can arise when \textit{time-dependent} controllers are applied to flows. 

The paper is organized as follows.  In Section~\ref{IHS} we will begin with some general remarks on integrable Hamiltonian systems and present our control law of interest.  In Section~\ref{1DHS} we will study our control law when applied to an integrable Hamiltonian system with one degree of freedom.  We will see that the degenerate shear flow dynamics inherent to an integrable Hamiltonian system will cause a \textit{Takens-Bogdanov} bifurcation when control is applied.  In Section~\ref{takens} we will discuss some aspects of the theory of Takens-Bogdanov bifurcations and their implications for robust control of the system.   Section~\ref{NLSsec} contains a detailed example which uses a driven NLS to illustrate the results of Sections~\ref{1DHS} and \ref{takens}.  Finally, in Section~\ref{concl} we will conclude with some ideas for future work in the area.

\section{Integrable Hamiltonian Systems and Control} \label{IHS}
Consider the following system of ordinary differential equations (ODEs):
\begin{equation} \label{openloop}
\dot {\mathbf{z}} = \mathbf{F}(\mathbf{z}),
\end{equation}
where $\mathbf{z},\mathbf{F} \in \mathbb{R}^{2N}$ and $\mathbf{z}$ are the ``lab" coordinates, understood as the ``natural" physical coordinates. The system (\ref{openloop}) is known as the ``open-loop" (or undriven/uncontrolled) dynamics.

In this paper, we specialize to $\mathbf{F}(\mathbf{z}) = J \nabla H$ where
\begin{equation}
J=
\left (
\begin{array}{cc}
0 & 1 \\
-1 & 0
\end{array}
\right ),
\end{equation}
and the 0's and 1's are $N\times N$ zero and identity matrices and $H(\mathbf{z})$ is the Hamiltonian, a scalar function of $\mathbf{z}$.  The system (\ref{openloop}) is called \textit{integrable} if there exists $N$ first integrals of (\ref{openloop}) which are independent and in involution.  If the  level sets of the integrals are compact, then regions of the phase space  are locally foliated by invariant manifolds with the topology of $N$-tori~\cite{Arnold}.  In what follows, we assume (\ref{openloop}) is an integrable Hamiltonian system and we study the control of (\ref{openloop}) on a typical (though arbitrary) invariant torus in its phase space.  

Consider a particular solution of (\ref{openloop}), $\mathbf{z_0}(t)$, which lies on one of those tori ($\dot{\mathbf{z}}_0=\mathbf{F}(\mathbf{z}_0)$).  We now apply a controller which targets $\mathbf{z_0}(t)$:
\begin{equation} \label{closedloop1}
\dot {\mathbf{z}} = \mathbf{F}(\mathbf{z}) +\epsilon \cdot \mathbf{f}(\mathbf{z_0},\mathbf{z}),
\end{equation}
where $\mathbf{z_0}(t) \in \mathbb{R}^{2N}$ and $\mathbf{f}$ is a $2N$-dimensional vector such that $\mathbf{f}(\mathbf{z_0},\mathbf{z_0})=0$.  The control coupling (gain), $\epsilon$, is a $2N\times 2N$ matrix whose entries need not be small. In principle, the control
law $\mathbf{f}$ can also involve the past history of $\mathbf{z}(t)$
(\textit{i.e.} `feedback').   The equation (\ref{closedloop1}) will be called our ``closed-loop" (or driven/controlled) dynamics.  Note that the control is applied in the ``physical coordinates", $\mathbf{z}(t)$.

Our problem is: How do we choose
$\mathbf{f}$ so that the given ``target" orbit, $\mathbf{z}_0$, in the open-loop dynamics becomes
an attractor in the closed-loop dynamics?  By adding the controller, $\mathbf{f}$, we are locally breaking up the tori and stabilizing one particular orbit. 

One choice for $\mathbf{f}$ is simply:
\begin{equation} \label{closedloop}
\dot {\mathbf{z}} = \mathbf{F}(\mathbf{z}) + \epsilon \cdot  (\mathbf{z_0}(t)-\mathbf{z}),
\end{equation}
where $\epsilon$ is the real $2N \times 2N$ matrix, $\epsilon=\epsilon_R 1 + \epsilon_I  J$, `$1$' is the $2N \times 2N$ identity matrix, and $(\epsilon_R,\epsilon_I)$ are real constants with $\epsilon_R>0$. 
As we we'll show, this form of control can, for large enough $\epsilon_R$, lead to \textit{synchronization} of $\mathbf{z}$ to our target orbit, $\mathbf{z_0}$~\cite{reggie}.  Notice,
\begin{equation}
\epsilon^2 = (\epsilon_R^2-\epsilon_I^2)1+2\epsilon_I \epsilon_R J,
\end{equation}
therefore $(\epsilon_R,\epsilon_I)$ act like real and imaginary parts of a complex scalar gain under matrix multiplication of $\epsilon$.

We can study the nature of the control law (\ref{closedloop}) by performing a linear analysis about the target, $\mathbf{z_0}(t)$.  Suppose $\mathbf{z}=\mathbf{z_0}(t)+\mathbf{\delta z}(t)$ and insert this into our closed-loop dynamics (\ref{closedloop}),
\begin{equation} \label{stability}
\dot {\mathbf{\delta z}}= J S(t) \mathbf{\delta z} -\epsilon_R \mathbf{\delta z},
\end{equation}
where $S(t)$ is a symmetric $2N \times 2N$ matrix and involves the Hessian of the Hamiltonian evaluated on the target orbit $z_0$,
\begin{equation} \label{s}
S_{jk}(t)=\frac {\partial^2 H}{\partial z_j \partial z_k} |_{\mathbf{z}=\mathbf{z_0}(t)} - \epsilon_I \delta_{jk},
\end{equation}
and $\delta_{jk}$ is the $2N \times 2N$ Kronecker delta function.  In the case of $\epsilon_R \uparrow \infty$ and $\epsilon_I=0$, (\ref{stability}) becomes $\dot {\mathbf{\delta z}} \approx -\epsilon_R \mathbf{\delta z}$.  This gives $\mathbf{\delta z}(t) \approx e^{-\epsilon_R t} \mathbf{\delta z}(0)$ which shows that in the limit of  $\epsilon_I=0$ the control law in (\ref{closedloop}) is \textit{purely dissipative} and $\mathbf{z}(t) \rightarrow \mathbf{z}_0(t)$ on a timescale of $O({\epsilon_R}^{-1})$.

Now consider the case where $\epsilon_R=0$.  For short times (\textit{i.e.} $t \rightarrow t+h$) $S(t)$ can be considered as a constant matrix and (\ref{stability}) integrates to,
\begin{equation} 
\mathbf{\delta z}(t+h) = \exp \left ( hJS \right ) \mathbf{\delta z}(t).
\end{equation}
It is known from the theory of  Lie groups~\cite{Arnold} that the matrix $M(t+h,t)\equiv \exp \left ( hJS(t,\epsilon_I) \right )$ is a symplectic matrix as long as $S$ is symmetric, which it is by (\ref{s}).  Hence, the control law in (\ref{closedloop}) generates symplectic maps in the case of $\epsilon_R = 0$ (\textit{i.e.} it generates time dependent canonical transformations).  The orbits near $\mathbf{z}_0$ neither attract to $\mathbf{z}_0$ nor repel away from it and the controller is \textit{conservative}.   

In Appendix~\ref{2ndapp}, we show that in the neighborhood of $\mathbf{z}_0(t)$ (\ref{closedloop}) can be rewritten in terms of a new set of canonical variables, $\mathbf{Z}$:
\begin{equation} \label{canonicalclosed}
\dot{\mathbf{Z}}=J\nabla_\mathbf{Z}K(\mathbf{Z},t)-\epsilon_R \mathbf{Z}-\epsilon_I JS_0(t)\mathbf{Z}+O(\mathbf{Z}^2),
\end{equation}
where $S_0(t)$ is a symmetric matrix and $K$ is a new Hamiltonian. 
In this paper we study the specific case in which $S_0(t)$ is constant.  The presence of non-constant $S_0(t)$ complicates the analysis and is beyond the scope of this paper.  As we will show, the presence of a constant $S_0(t)$ is already a serious complication in terms of nonlinear analysis of the control law.  In Section~\ref{NLSsec}, we will see that this simplification (of a constant $S_0(t)$) holds for the nonlinear Schr\"{o}dinger equation.

The goal of our work is to turn the target orbit $\mathbf{z}_0(t)$ into an attractor.  We wish to understand the geometry of the attractor basin and the topology of nearby orbits.   Doyon and Dub\'{e}~\cite{doyon} demonstrate targeting 
periodic orbits of a particular period, $m$,  in Hamiltonian systems when the location and the stability are unknown and the dynamics of the system are chaotic.  Our work complements this result in that we will be studying the consequences of using both dissipative and conservative control for an \textit{integrable} Hamiltonian system onto a \textit{known} orbit.  In addition, we focus on the local question of the closest `distance' to the basin boundary of our new attractor.  This distance strongly depends upon whether the control is conservative or dissipative.  Note that there is no true meaning of `distance' in phase space, hence by that term we mean the typical noise level which would destabilize the target.  We will show that, due to the shear flow structure inherent to integrable Hamiltonian systems, something known as a ``Takens-Bogdanov'' bifurcation generically occurs when control is applied.  As we will see, the appearance of a Takens-Bogdanov bifurcation implies that the evolution of the system will be very sensitive to noise and parameter uncertainty~\cite{TB1,TB2} in the purely dissipative limit of (\ref{closedloop}).  We will also see that by turning on the conservative part of the drive, $\epsilon_I$, the controllability is improved.

It should be pointed out that Haberman and Ho~\cite{haberman} have studied dissipatively perturbed Hamiltonian systems in a  regime which contains two competing centers (which become attractors once the dissipation has been ``turned on'') separated by a saddle in the phase space before the perturbation is applied.  Their Hamiltonian system is a nonlinear oscillator where the drive frequency is near (or at) the natural frequency of the system.  Although the phase space topology they study is similar to ours, we are asking different questions.  Their work is concerned with deriving an analytic form for the stable manifold of the saddle (basin boundary) using asymptotic methods (once small dissipation has been applied). The work presented here is concerned with the generic properties of both dissipative and conservative control laws applied to a general integrable Hamiltonian system.    

\section{Hamiltonian Systems with One Degree of Freedom} \label{1DHS}
\begin{figure}[ht]
\begin{center}
\scalebox{0.4}{{\includegraphics{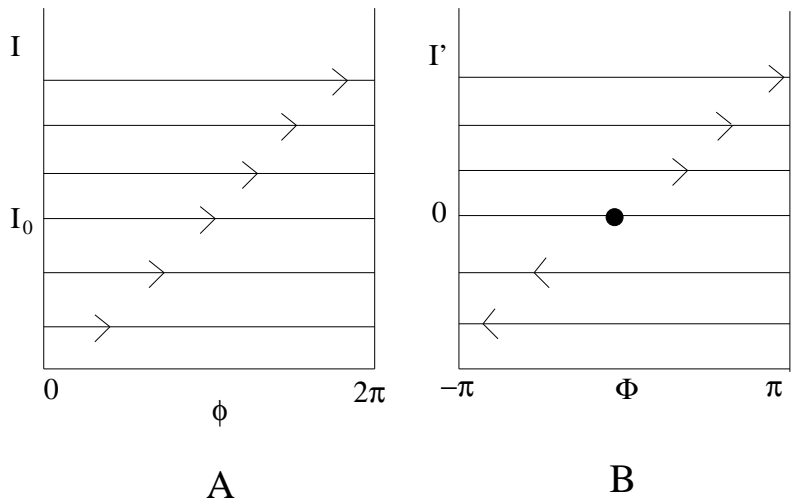}}}
\end{center}
\caption{Figure~\ref{figure:AAdynamics}A shows the flow field of the open-loop dynamics of (\ref{AA}).  Figure~\ref{figure:AAdynamics}B shows the flow field of the open-loop dynamics of the transformed coordinate system (\ref{gen}).  Notice that the linear part of (\ref{gen}) has only one eigenvector which lies along the line of stagnation ($I'=0$).}
\label{figure:AAdynamics}
\end{figure}

Consider a Hamiltonian system with one degree of freedom and Hamiltonian $H=H(q,p)$, with $q,p \in \mathbb{R}$.  The evolution of $p$ and $q$ is dictated by the canonical equations:
\begin{eqnarray} \label{pq}
\dot q&=&\frac{\partial H(p, q)}{\partial p}, \nonumber\\
\dot p&=&-\frac{\partial H( p, q)}{\partial q}. 
\end{eqnarray}
Suppose that for this system, the Hamiltonian has regions with compact level sets, implying there are regions of the phase space which are foliated by circles (1-tori)~\cite{Arnold}.  These circles are invariant under the flow generated by $H(p,q)$.  On a given family of these tori, the coordinates $(p,q)$ can be canonically transformed to the action-angle coordinates $(I,\phi)$ and the Hamiltonian can be written as $H(p,q)=H(I)$. The evolution of $(I,\phi)$ is of the form:
\begin{eqnarray} \label{AA}
\dot I & = & 0, \nonumber\\
\dot \phi & = & \frac{\partial H(I)}{\partial I} \equiv \omega(I).
\end{eqnarray} 
In these coordinates the dynamics of (\ref{AA}) looks locally like a shear flow with each neighboring torus having a slightly different (constant) rotation rate (see Figure~\ref{figure:AAdynamics}).  The evolution of the action-angle variables is quite simple, making them the natural coordinates for this region of the phase space. It is important to keep in mind, however, that the evolution in the original $(p,q)$ coordinates, although periodic, can be quite complicated.   

The system (\ref{pq}), or equivalently (\ref{AA}), will be our open-loop dynamics for control.  The technique used to control (\ref{pq}) is simple.  First we must choose some target orbit of (\ref{pq}), $(I_0,\phi_0)$, where 
\begin{equation}
\phi_0(t)=\omega_0 t + \delta
\end{equation}
with $\delta$ being an arbitrary angle between 0 and $2\pi$ and $\omega_0=\partial H(I)/ \partial I |_{I_0}$.   Without loss of generality, we will set $\delta=0$.  After choosing a target, we transform into a canonical coordinate system where the target is fixed at the origin, using a time dependent canonical transformation.  This puts the target orbit at rest at the origin.  We then turn on the controller which converts the origin into an attractor.  We will find that applying the controller results in what is known as a Takens-Bogdanov bifurcation~\cite{guckenheimer}.  The presence of the Takens-Bogdanov bifurcation has an important effect on our ability to control the system to the target.  Further, we'll see that as the control becomes more conservative (\textit{i.e.} as $\epsilon_I/\epsilon_R \uparrow \infty$), the controllability, as measured by the minimum `kick' required to destabilize the target, is greatly improved even when ${\epsilon_R}^2+{\epsilon_I}^2$ is fixed.  However, there is a tradeoff:  while a large $\epsilon_I$ and small $\epsilon_R$ may produce a large basin of attraction it will also have a long decay time to the target.

Let's choose our target to be: $I=I_0$ and $\phi=\phi_0(t)$ with $\phi_0$ defined as above.  We perform the previously mentioned canonical coordinate transformation using the generating function~\cite{percival,goldstein} $F_2( I',\phi,t)=(\phi-\omega_0 t)( I'+I_0)$ with:
\begin{eqnarray}
I & = & \frac{\partial F_2}{\partial \phi}= I' + I_0, \nonumber \\
\Phi & = & \frac{\partial F_2}{\partial I'}=\phi-\omega_0 t, \nonumber \\
K( I')&=&H( I'+I_0)+\frac{\partial F_2}{\partial t}=H( I'+I_0)-  (I'+I_0)\omega_0. \label{boost}
\end{eqnarray}
Note that this transformation places the control target at the origin, $(I',\Phi)=(0,0)$.  

Now we examine the dynamics about the origin via Taylor exapansion of the new Hamiltonian, $K(I')$ about the target:
\begin{equation}
K(I') = H(I_0) + \frac{\partial H}{\partial I'}|_{I_0}I' + \frac{1}{2}\frac{\partial^2 H}{\partial I'^2}|_{I_0}{I'}^2 + \cdots -(I_0+I')\omega_0.
\end{equation}
Next, we ignore the constant terms, $H(I_0)$ and $\omega_0I_0$, and collect terms of $O({I'}^3)$ and higher into a function $h( I')$,
\begin{equation}
K( I')=\frac{\lambda}{2} { I'}^2+h(I').
\end{equation}
This Hamiltonian gives the following equations of motion,
\begin{equation} \label{gen}
\left (
\begin{array}{c}
\dot {\Phi} \\
\dot { I'} 
\end{array}
\right ) =
\left (
\begin{array}{cc}
0 & \lambda \\
0 & 0
\end{array}
\right )
\left (
\begin{array}{c}
{\Phi} \\
{ I'} 
\end{array}
\right ) +
\left (
\begin{array}{c}
f(I') \\
0
\end{array}
\right ),
\end{equation}
where $f(I')=dh/dI'$ is $O({I'}^2)$.  The equation (\ref{gen}) is our open-loop dynamics and describes a shear flow with the entire $I'=0$ line fixed.  Figure~\ref{figure:AAdynamics} shows the flow field of the dynamics of (\ref{AA}) and (\ref{gen}). Such shear flow dynamics, characterized by a degenerate linear term, are the setting for a Takens-Bogdanov bifurcation~\cite{TB2}. 

Next, we examine the closed-loop dynamics.  Recall equation (\ref{closedloop}):
\begin{equation} 
\dot{\mathbf{z}} = \mathbf{F}(\mathbf{z}) +(\epsilon_R1+ \epsilon_I J)(\mathbf{z_0}(t)-\mathbf{z}), \nonumber
\end{equation}
with $\mathbf{F}(\mathbf{z})=J\nabla H(\mathbf{z})$ and $\mathbf{z_0}(t)$ is a solution to the open-loop dynamics ($\epsilon_R=\epsilon_I=0$).  While the second term of the RHS of (\ref{closedloop}) is linear in the ``physical coordinates" $z=(p,q)$ it will, in general, contain \textit{nonlinear} terms in $z=(I',\Phi)$ due to the nonlinear nature of the transformation from $(p,q)$ to $(I',\Phi)$.  Using (\ref{gen}) and the results from Appendix~\ref{2ndapp}, we can always write the dynamics in the following form (possibly after a near identity canonical transformation on ($I',\Phi$)):
\begin{equation} \label{genericcontrol}
\left (
\begin{array}{c}
\dot {\Phi} \\
\dot{I'} 
\end{array}
\right ) =
\left (
\begin{array}{cc}
-\epsilon_R & \lambda \\
-\eta \epsilon_I & -\epsilon_R
\end{array}
\right )
\left (
\begin{array}{c}
\Phi  \\
I'
\end{array}
\right ) +
\left (
\begin{array}{c}
f_1(I',\Phi;\epsilon) \\
f_2(I',\Phi;\epsilon)
\end{array}
\right ),
\end{equation}
where the $\epsilon$ notation in $f_i$ refers to both $\epsilon_R$ and $\epsilon_I$.  The minus sign in front of the $\epsilon_I$ term ensures that a center opens up at the target (when the product $\lambda \eta \epsilon_I >0$), as will be discussed below, and:
\begin{eqnarray}
f_1(I',\Phi;\epsilon)  &=& f(I') + \epsilon \tilde{f}_1(I',\Phi;\epsilon), \nonumber \\
f_2(I',\Phi;\epsilon)  &=& \epsilon \tilde{f}_2(I',\Phi;\epsilon).  
\end{eqnarray}
The general solution to the \textit{linear} dynamics of (\ref{genericcontrol}) is:
\begin{equation}\label{exactsoln}
\left (
\begin{array}{c}
\Phi(t) \\
I'(t) 
\end{array}
\right ) =
e^{-\epsilon_Rt}
\left (
\begin{array}{cc}
\cos(\sqrt{\lambda \eta \epsilon_I}t)  & \sqrt{\frac {\lambda}{\eta \epsilon_I}} \sin(\sqrt{\lambda \eta \epsilon_I}t) \\
-\sqrt{\frac {\eta \epsilon_I}{\lambda}} \sin(\sqrt{\lambda \eta \epsilon_I}t) & \cos(\sqrt{\lambda \eta \epsilon_I}t)
\end{array}
\right )
\left (
\begin{array}{c}
\Phi (0)  \\
I' (0)
\end{array}
\right ). 
\end{equation}
Hence, we can see that the system undergoes a decaying oscillation which will eventually settle onto the target (origin) with a decay time scale of $O(1/ \epsilon_R)$ and oscillation period $O(1/ \sqrt{\epsilon_I})$.

The nonlinear behavior of systems with linear degeneracy can be subtle.
To develop an idea of how each term in the closed-loop dynamics (purely dissipative and conservative) behaves, we consider two limits:  \romannumeral 1) $\epsilon_R \ne 0$, $\epsilon_I=0$, and \romannumeral 2) $\epsilon_I \ne 0$, $\epsilon_R=0$.  We first treat their linear behavior.  

In Case \romannumeral 1) (\ref{closedloop}) becomes,
\begin{equation}
\dot{\mathbf{z}} = \mathbf{F}(\mathbf{z}) +\epsilon_R(\mathbf{z_0}(t)-\mathbf{z}), \nonumber
\end{equation}
which is a purely dissipative drive, therefore we expect both linear and nonlinear dissipative terms in $(I',\Phi)$  coordinates.  In action-angle coordinates the dynamics becomes (see Appendix~\ref{2ndapp}), 
\begin{equation} \label{genericdiss}
\left (
\begin{array}{c}
\dot {\Phi} \\
\dot{I'} 
\end{array}
\right ) =
\left (
\begin{array}{cc}
-\epsilon_R & \lambda \\
0 & -\epsilon_R
\end{array}
\right )
\left (
\begin{array}{c}
\Phi \\
 I' 
\end{array}
\right ) +
\left (
\begin{array}{c}
f_1(I',\Phi,\epsilon_R) \\
f_2(I',\Phi,\epsilon_R)
\end{array}
\right ).
\end{equation}
In general, $f_1$ and $f_2$ will be nonlinear functions in $I'$ and $\Phi$ and contains both linear and nonlinear terms in $\epsilon_R$.  It is clear that
the linearized dynamics of (\ref{genericdiss}) is non-diagonalizable even though linearly stable.  A Takens-Bogdanov bifurcation can occur and, as we will see in Section~\ref{takens}, will limit our ability to control (\ref{pq}) to our desired state.  In Section~\ref{takens} we will also discuss the effect of breaking the degeneracy in the diagonal term of(\ref{genericdiss}).

The solution for the linear dynamics of (\ref{genericdiss}) is:
\begin{equation}
\left (
\begin{array}{c}
\Phi (t) \\
I' (t) 
\end{array}
\right ) =
\left (
\begin{array}{cc}
e^{-\epsilon_Rt} & \lambda t e^{-\epsilon_Rt} \\
0 & e^{-\epsilon_Rt}
\end{array}
\right )
\left (
\begin{array}{c}
\Phi (0) \\
 I'(0) 
\end{array}
\right ).
\end{equation}
Where the $t\exp(-\epsilon_Rt)$ term represents an effect known as \textit{transient amplification} which will be described in Section~\ref{takens}.  After the transient amplification, we see exponential decay to the target on a time scale of $1/\epsilon_R$.

\begin{figure}[ht]
\begin{center}
\scalebox{0.4}{{\includegraphics{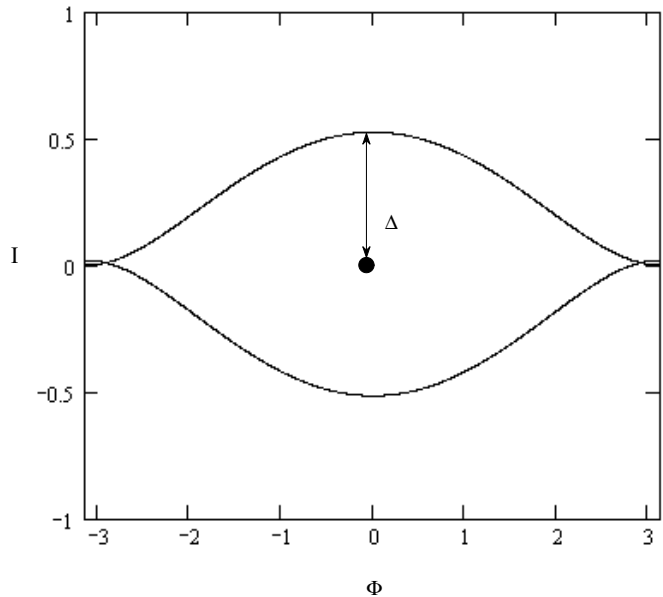}}}
\end{center}
\caption{The island opening around the target at the origin is due to the presence of a conservative controller in resonance with the target solution (the origin).  Here $\epsilon_I=0.4$ and $\eta=1$.  Notice the large width of the island, $\Delta=O(\sqrt{\epsilon_I}$).}
\label{figure:island}
\end{figure}

In Case \romannumeral 2) (\ref{closedloop}) becomes,
\begin{equation} \label{conservagain}
\dot{\mathbf{ z}} = \mathbf{F}(\mathbf{z}) +\epsilon_I J \cdot (\mathbf{z_0}(t)-\mathbf{z}), \nonumber 
\end{equation}
which in Section~\ref{IHS} was shown to be a conservative drive.  When the drive is periodic and resonant, we expect islands to open in the phase space centered around our target orbit.  It is well known (see Appendix~\ref{1stapp}) that the width of those islands will generically be of $O(\sqrt{\epsilon_I})$.  Figure~\ref{figure:island} illustrates this.

When $\epsilon_R=0$ (the \textit{purely conservative case}), the equation (\ref{genericcontrol}) becomes:
\begin{equation} \label{genericconserv}
\left (
\begin{array}{c}
\dot {\Phi}  \\
\dot I' 
\end{array}
\right ) =
\left (
\begin{array}{cc}
0 & \lambda \\
-\eta \epsilon_I & 0
\end{array}
\right )
\left (
\begin{array}{c}
\Phi  \\
I' 
\end{array}
\right ) +
\left (
\begin{array}{c}
f_1(I',\Phi ,\epsilon_I) \\
f_2(I',\Phi ,\epsilon_I)
\end{array}
\right ),
\end{equation}
where $\eta$ is constant, and $f_i$ are nonlinear functions of $(I',\Phi)$, and in geneneral, contain linear and nonlinear terms in $\epsilon_I$ and will be zero at the origin (note that the $f_i$'s will be different than those in (\ref{genericcontrol}) and (\ref{genericdiss})).  Because we have applied a conservative controller to our system, the closed-loop dynamics (\ref{genericconserv}) is also a Hamiltonian system;  therefore areas in the phase space are conserved, and no attractors can be present.  In terms of control, this means that if the system is in an initial state which is inside the island, then the trajectory ``orbits" the target but never settles onto it.  Some dissipation must be present in order for control to be achieved.  

Setting $\epsilon_R=0$ in (\ref{exactsoln}) the linear term of (\ref{genericconserv}) can be solved to give, 
\begin{equation} \label{linconserv}
\left (
\begin{array}{c}
\Phi(t) \\
I'(t) 
\end{array}
\right ) =
\left (
\begin{array}{cc}
\cos(\sqrt{\lambda \eta \epsilon_I}t)  & \sqrt{\frac {\lambda}{\eta \epsilon_I}} \sin(\sqrt{\lambda \eta \epsilon_I}t) \\
-\sqrt{\frac {\eta \epsilon_I}{\lambda}} \sin(\sqrt{\lambda \eta \epsilon_I}t) & \cos(\sqrt{\lambda \eta \epsilon_I}t)
\end{array}
\right )
\left (
\begin{array}{c}
\Phi (0)  \\
I' (0)
\end{array}
\right ). 
\end{equation}
Hence, we see that the trajectories orbit the target with frequency $\sqrt{\lambda \eta \epsilon_I}$ and never settle on to it.  Further, the $ \sqrt{\frac {\lambda}{\eta \epsilon_I}}$ term in the off-diagonal entries of (\ref{linconserv}) show that the island which is formed is very thin.  This can be seen by following an orbit which starts at $(I'(0)=0,\Phi (0)=\Delta \Phi)$.  The orbit, centered about the origin, is elliptical and will intersect the $I'$ axis when $t=t_1=\frac{\pi}{2\sqrt{\lambda \eta \epsilon_I}}$ (one quarter of the period).  We can see from (\ref{linconserv}), that the orbit intersects the $I'$ axis at $(I'=-\sqrt{\frac{\eta \epsilon_I}{\lambda}}\Delta \Phi,\Phi=0)$, hence the island is very thin $(O(\sqrt{\epsilon_I}))$ compared to the initial $\Delta \Phi$ displacement.  However, this island will `open up' rapidly as $\epsilon_I$ is increased.        

In the next section, we will show that the presence of the $\epsilon_I$ terms in the linearized dynamics enlarges the basin of attraction associated with the Takens-Bogdanov bifurcation and thus improves the control.  In Section~\ref{NLSsec} we will give an explicit example of this result using a driven NLS equation.

\section{Takens-Bogdanov Bifurcations} \label{takens}

In Section~\ref{1DHS}, we showed that, in general, the linearized dynamics of an integrable Hamiltonian system becomes 
non-diagonalizable in the limit of no control ($\epsilon_R \downarrow 0$ and $\epsilon_I \rightarrow 0$).  When a system's linearized dynamics becomes non-diagonalizable, a Takens-Bogdanov bifurcation occurs.   The interested reader is directed to the most recent edition of~\cite{guckenheimer} for a thorough discussion of Takens-Bogdanov bifurcations and~\cite{TB1,TB2} for a discussion of estimating the distance to the basin boundary in the subcritical case, and aspects of noise driven escape in the purely dissipative case.  We do not consider the full `unfolding' of the Takens-Bogdanov bifurcation, but only those parameter ranges relevant to the present control problem (\textit{i.e.} those having a basin of attraction).  In particular, we study the unfolding using the natural parameterization inherited from of our control law rather than that of ~\cite{guckenheimer} (note that we keep the target fixed at the origin, while the standard parameterization moves it).  In this section, we will present the information most relevant for our work here as well as expand upon the results of~\cite{TB1,TB2} to include the conservative term.

We begin by considering the purely dissipative case, $\epsilon_I=0$.  We will use (\ref{genericdiss}) as our generic example of a system exhibiting a Takens-Bogdanov birfurcation.  For the situation of interest here, the Takens-Bogdanov 
bifurcation involves a degenerate saddle-node bifurcation in (\ref{genericdiss}) when $\epsilon_R \downarrow 0$.  The degenerate node at the origin will be the solution of (\ref{pq}) which is the target for control.

\begin{figure}[ht]
\begin{center}
\scalebox{0.4}{{\includegraphics{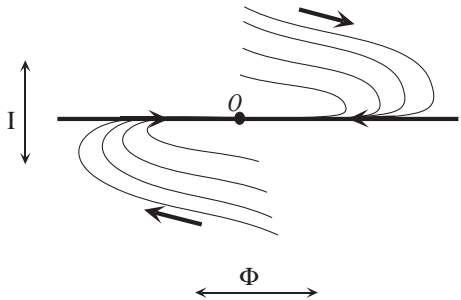}}}
\end{center}
\caption{
A qualitative sketch of the dynamics when $\epsilon_R>0$.  The node is denoted by the point, $O$.  The nonlinear terms have not yet been added.  The effect of ``transient amplification" is clearly present.  The horizontal line is the old line of neutral stability which separated the two shear flow regions.}
\label{figure:node}
\end{figure}

When $\epsilon_R=0$ (Figure~\ref{figure:AAdynamics}B), we have a line of neutral stability which bounds two shear flow regions ($I'>0$, and $I'<0$).  When $\epsilon_R$ becomes greater than zero (Figure~\ref{figure:node}) a stable node appears, denoted $O$.  
The effect of the previous shear flow can still be seen, however, as the trajectories must approach the node tangentially along the old neutrally stable line with very slow transverse dynamics.  The further a trajectory starts away from the node the greater the effect of the shear flow which forces the trajectory to travel further in the horizontal direction before being attracted to the node.   This effect is known as ``transient amplification" since the distance to the node will typically grow, before the slow decay to the node sets in.  If the diagonal terms of (\ref{genericdiss}) are slighly different ($\epsilon_{R1} \ne \epsilon_{R2}$), Figure~\ref{figure:node} changes slightly because the exact degeneracy of the eigenvectors is lifted.  However, transient amplification still occurs since it is due to the linear term being \textit{ill-conditioned}~\cite{TB1,trefethen,farrell1,farrell2}.

\begin{figure}[ht]
\begin{center}
\scalebox{0.4}{{\includegraphics{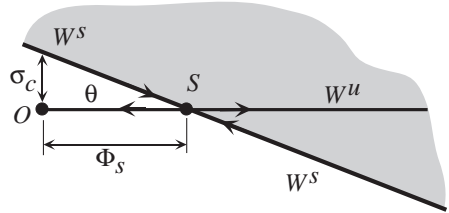}}}
\end{center}
\caption{
Illustration of the ``triangle relation" of~\cite{TB2}.  The shortest distance to the basin boundary, $\sigma_c$, is now along some other direction in the phase space besides the saddle-sink	connection.}
\label{figure:triangle}
\end{figure}

\begin{figure}[ht]
\begin{center}
\scalebox{0.4}{{\includegraphics{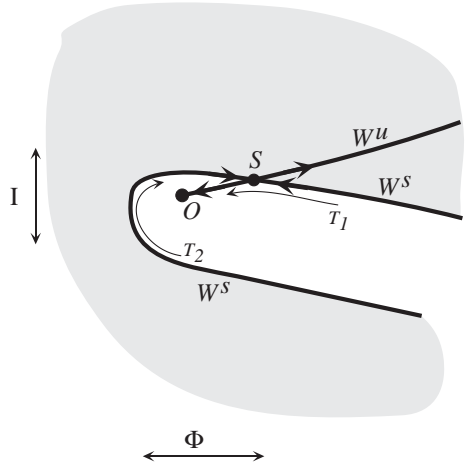}}}
\end{center}
\caption{
A typical basin of attraction for a node undergoing a Takens-Bogdanov bifurcation. Notice the distinctive tear drop shape of the basin due to the near-degeneracy of $W^s$ and $W^u$.}
\label{figure:wholebasin}
\end{figure}

 Provided certain conditions are met (described in~\cite{TB2}) a saddle-point appears in the neighborhood of the node.  In Figure~\ref{figure:triangle}, we zoom in on the area around the node with $\epsilon_R>0$ and we include the nonlinear terms $f_1$ and $f_2$.  
One half of the 
line of neutral stability becomes one half of the unstable manifold ($W^u$) for the saddle point, sometimes called the \textit{saddle-sink connection}.  This piece of the unstable manifold ends at the degenerate node, $O$.  The other piece of the unstable manifold ($W^u$) can lead to another attractor of some type (not shown).  In Figure~\ref{figure:wholebasin}, we show the full basin of attraction for the node, $O$.  
The saddle point, $S$, also has a stable manifold ($W^s$) which forms the boundary of the basin of attraction for $O$. Extending $W^s$ typically gives a tear-drop shaped basin near $O$  which is one hallmark of a Takens-Bogdanov bifurcation.  Some sample trajectories are included in Figure~\ref{figure:wholebasin} and labeled, $T_i$.  Notice that while one of the trajectories, $T_1$, is attracted directly into the node, the other trajectory, $T_2$, misses the node on its first approach and must travel in the vicinty of $W^s$ before connecting to the node.  There is only one direction in which a trajectory may approach the node, along the saddle-sink connection.  The tear-drop shape for the basin comes about because, as $\epsilon_R \downarrow 0$, the eigenvectors of the linear dynamics in the vicinity of the saddle become degenerate (parallel) and, therefore, the angle between the stable and unstable manifold of $S$ decreases.  The near-degeneracy of $W^u$ and $W^s$ can be seen in Figure~\ref{figure:wholebasin}.   

Due to the shape of the basin of attraction, systems exhibiting a Takens-Bogdanov bifurcation are extremely sensitive to 
noise and parameter uncertainty~\cite{TB1}, and the subcritical threshold of instability (\textit{i.e.} the distance to the basin boundary) scales differently from normal saddle-node and Hopf bifurcations~\cite{TB2}.  Both~\cite{TB1} and~\cite{TB2} demonstrate that the distance to the basin boundary, $\sigma_c$, is proportional to ${\epsilon_R}^\gamma$ ($\gamma>1$), where $\gamma$ can be computed using a simple formula once the normal form of the dynamics in the neighborhood of the target is found.  A normal form analysis simplifies the dynamical system near the target using near-identity transformations.  The normal form reveals which nonlinear terms govern the topology of the phase space near the target (\textit{e.g.} the location of the basin boundary of the target) and, for $\epsilon_R \gg \epsilon_I$ (see below), gives the location of the saddle, $S$.  In some applications, noise may be applied randomly in the phase space, and therefore $\sigma_c$ also gives a noise threshold for instability.  However, when $\epsilon_R \downarrow 0$ and $\epsilon_I \ne 0$, the normal form can give misleading results, as will be shown.

We consider the case in which the controller contains both dissipative and conservative terms.  Our goal is to derive a  subcritical threshold scaling, $\sigma_c(\epsilon_R,\epsilon_I)$.  We will use a method similar to that in~\cite{TB2}.  Following our results from Section~\ref{1DHS}, the normal form of our model system to leading order in the vicinity of the control target is:
\begin{equation} \label{conservmodel}
\left (
\begin{array}{c}
\dot \Phi \\
\dot I' 
\end{array}
\right ) =
\left (
\begin{array}{cc}
-\epsilon_R & \lambda \\
-\eta \epsilon_I & -\epsilon_R
\end{array}
\right )
\left (
\begin{array}{c}
\Phi \\
I'
\end{array}
\right ) +
\left (
\begin{array}{c}
0 \\
b_1 \Phi^2 + b_2 \Phi I'
\end{array}
\right ),
\end{equation}
where we have chosen $\lambda \eta \epsilon_I>0$ and $b_1,b_2$ are constants.  The open-loop ($\epsilon_R=\epsilon_I=0$) form of (\ref{conservmodel}) is sometimes known as the \textit{Bogdanov form}~\cite{guckenheimer}.  The nonlinear terms are the generic dominant terms for a Takens-Bogdanov system~\cite{TB2}, found by casting the open-loop dynamics of (\ref{genericcontrol}) into normal form.  We now describe how to estimate $\sigma_c$ using the triangle relation of~\cite{TB2}. 

First, find the location of the saddle by setting $\dot \Phi = \dot I = 0$ and solving:
\begin{equation} \label{saddleloc}
 \Phi_S = \frac{\epsilon_R^2+\lambda \eta \epsilon_I}{(b_1\lambda+\epsilon_R b_2)},\mbox{ } {I'}_S=\frac{(\epsilon_R^2+\lambda \eta \epsilon_I)\epsilon_R}{(\lambda b_1+\epsilon_R b_2)\lambda }.
\end{equation}  
As in~\cite{TB2}, we see that the $\Phi^2$ term is dominant if $\lambda b_1$ and $b_2$ are $O(1)$.  It is important to note, however, that when $\epsilon_R=0$ and $\epsilon_I \ne 0$, the dynamics of (\ref{conservmodel}) are diagonalizable and \textit{not} degenerate. In this case, the normal form is suspect because we need to include higher order terms so that we may correctly describe the location of the saddle and the shape of $W^s$.  In what follows, we assume $\epsilon_I \ll \epsilon_R$.  

Second, linearize (\ref{conservmodel}) about the saddle point ($\Phi=\Phi_S+u,I={I}_S+v$):
\begin{equation} \label{saddyn}
\left (
\begin{array}{c}
\dot u \\
\dot v 
\end{array}
\right ) =
\left (
\begin{array}{cc}
-\epsilon_R & \lambda \\
\frac{2{\epsilon_R}^2}{\lambda}+\eta \epsilon_I & -\epsilon_R
\end{array}
\right )
\left (
\begin{array}{c}
u \\
v
\end{array}
\right ) +
\left (
\begin{array}{c}
0 \\
b_1 u^2
\end{array}
\right ).
\end{equation}
The eigenvalues of the linear dynamics are:  $-\epsilon_R \pm \sqrt{2{\epsilon_R}^2+\lambda \eta \epsilon_I}$.  This verifies that the second fixed point is indeed a saddle-point (recall, we fix $\eta \epsilon_I >0$), for $\lambda >0$.  

Third, find the angle between $W^s$ and $W^u$ in Figure~\ref{figure:triangle}, which we denote, $\theta$.  This is done by finding the eigenvectors of the linear dynamics of (\ref{saddyn}) and using the cosine relation for the dot product of the two vectors.  It can be shown that:
\begin{equation}
\theta = \frac{\sqrt{2}}{\lambda}\sqrt{2\epsilon_R^2+\lambda\eta\epsilon_I}.
\end{equation}  

\begin{figure}[ht]
\begin{center}
\scalebox{0.4}{{\includegraphics{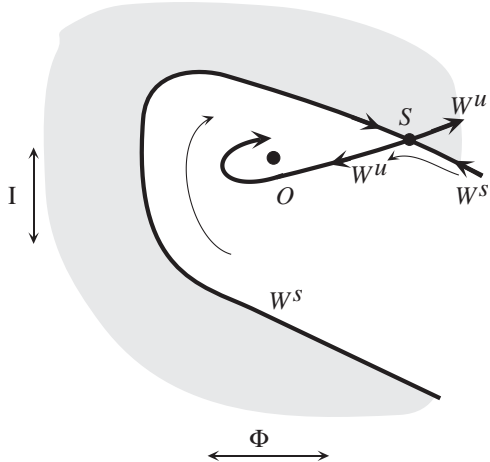}}}
\end{center}
\caption{
A qualitative sketch of a typical basin of attraction for a node in a Takens-Bogdanov system with a conservative term present.  Notice how much larger the basin is here, as compared to Figure~\ref{figure:wholebasin} and note also the target point is now a stable spiral.  }
\label{figure:TBconservbasin}
\end{figure}

Finally, the triangle relation from Figure~\ref{figure:triangle}~\cite{TB2} provides the estimate:
\begin{equation} \label{newscale1}
\sigma_c \approx \Phi_S\theta = \frac{\sqrt{2}}{\lambda}\frac{\epsilon_R^2+\lambda \eta \epsilon_I}{(b_1\lambda+\epsilon_R b_2)}\sqrt{2\epsilon_R^2+\lambda\eta\epsilon_I},
\end{equation}
where $\epsilon_I \ll \epsilon_R$.  For discussion of the higher order corrections to (\ref{newscale1}), see Appendix~\ref{appendix:error}.  As demonstarted in Appendix~\ref{appendix:error}, the largest source of error in the triangle relation comes from not including the curvature of the stable manifold in (\ref{newscale1}).  The triangle relation is the first order term in a Taylor series approximation of the stable manifold.  In Appendix~\ref{appendix:error}, we estimate the error in (\ref{newscale1}) by including the next higher order terms in the Taylor series approximation.  We see from (\ref{newscale1}) that in the purely dissipative case, $\epsilon_I=0$, $\sigma_c \approx {\epsilon_R}^3$, and we recover the result from~\cite{TB2}.  Hence, (\ref{newscale1}) shows that the presence of a conservative term in the control law increases $\sigma_c$.  This is sketched in Figure~\ref{figure:TBconservbasin}.  Notice that in Figure~\ref{figure:TBconservbasin} that $\sigma_c$ has dramatically increased as compared to Figure~\ref{figure:wholebasin} and now, the node has become a degenerate spiral as demonstrated in (\ref{exactsoln}).

In Section~\ref{NLSsec}, we will apply the results from this section and Section~\ref{1DHS} to a simple
control scheme for a driven Nonlinear Schr\"{o}dinger Equation.  

\section{Control of the Nonlinear Schr\"{o}dinger Equation} \label{NLSsec}

The one dimensional focusing nonlinear Schr\"{o}dinger Equation (NLS),
\begin{equation}\label{NLS}
iq_t+q_{xx}+2|q|^2q=0,
\end{equation}
 governs the 
envelope dynamics of waves that, to leading order, are weakly nonlinear, nearly monochromatic, and dispersive.  Here, we consider solutions of (\ref{NLS}) that are periodic in space (i.e. $q(x,t)=q(x+L,t)$ for some box size, $L$, and we choose $L=1$).  The NLS is used as a model system in many areas of physics such as plasmas, water waves, and nonlinear optics. 
The interested reader is referred to~\cite{GeneNLS} which contains many  references for applications of the NLS in its introduction and develops a special class of exact solutions to the NLS, specifically those associated with modulational instabilities. Ultimately, we wish to use this special class of solutions to design control laws for 
NLS-type systems such as the Ginzburg-Landau equation~\cite{GL} and Dysthe's equation~\cite{Dysthe}. The Ginzburg-Landau and Dysthe equations are not integrable, but NLS appears as a limiting case.  It is hoped that the current work described here will provide insights in developing control laws for these non-integrable systems as well.  Such control laws, for example, might be designed to suppress instabilites or to exploit them for pulse formation.  That work will be reported elsewhere.  In what follows (\ref{NLS}) will be our open-loop dynamics.

The NLS is an integrable Hamiltonian system with Hamiltonian, 
\begin{equation}\label{NLSham}
H=\int_{0}^1(|q|^4- |q_x|^2 ) dx.
\end{equation}
where $q^*$ is the complex conjugate of $q(x,t)$.  We restrict our 
attention to $q,q^*$ that are $C^\infty$.  We consider the set of all such functions to be our `phase space' with ($q,q^*$) as dynamical variables on that space.   The phase space variable, $q$, can be written as,
\begin{equation}\label{q}
q(x,t)=\sum_{n=-\infty}^\infty a_n(t)\exp(2\pi inx/L),
\end{equation}
thus each $e^{2\pi inx/L}$ define a `basis direction' in the phase space, which is infinite-dimensional and has embedded within it invariant $N$-tori.  The $\Theta$-function solutions of~\cite{GeneNLS} give explicit representations of the dynamics on the $N$-tori.  In what follows, we will be interested only in $q_0(x,t)=a\exp(2ia^2t)$, thus restricting ourselves to targets that lie on a 1-dimensional invariant torus.  This allows a very complete analysis and shows that this control problem is exactly of the form discussed in Section~\ref{1DHS}.

\subsection{The Control Law}

Our goal is to control (\ref{NLS}) to 
some target $q_0(x,t)$ which is an exact solution to (\ref{NLS}). 
We proceed as we did above by choosing our closed-loop dynamics to be:
\begin{equation} \label{closednls}
iq_t+q_{xx}+2|q|^2q=i(\epsilon_R+i\epsilon_I)(q_0-q),
\end{equation}
Note that the presence of the $i$ in front of the control law keeps $\epsilon_R$ the dissipative part of the control and $\epsilon_I$ the conservative part as was our convention in Section~\ref{IHS}.  This control law is the same as (\ref{closedloop}).  We choose $\epsilon_I<0$ so that the target, $q_0$, is at the center of the island in the conservative limit ($\epsilon_R=0$) as discussed in Section~\ref{1DHS}.  

Equation (\ref{closednls}) is a particular example of a driven nonlinear Schr\"{o}dinger equation and has been studied extensively by Li, McLaughlin \textit{et al}.~\cite{Li}, Haller~\cite{Haller}, and by Li and Wiggins~\cite{LiWiggins} in the case of $\epsilon_I=0$.  This body of work has revealed the rich geometrical structure  that exists in the solution space of the driven NLS.  These authors have extended the finite dimensional methods of invariant manifolds, Melnikov theory, homoclinic tangles, \textit{etc}. to the infinite-dimensional solution space of this nonlinear PDE.  We, however, will ask different questions.  As stated earlier, our ultimate interest is to learn how to control physical systems for which NLS-type  dynamics are reasonable models.  We will exploit the integrability of the open-loop dynamics to gain insight into geometrical aspects of the control problem.  The previously mentioned authors found that
complex behavior exists throughout the NLS's solution space. Our goal is to suppress this behavior in the neighborhood of certain target solutions.   

We consider `plane wave', \textit{i.e.} spatially uniform, solutions of (\ref{NLS}) hence, $q_{xx}=0$. Fix $q_0$ to be  
$q_0(t)=a\exp(2ia^2(t-t_0))$ where $a$ is some real positive constant. For simplicity, we will choose $t_0=0$.  Hence, we first restrict ourselves to the invariant manifold of plane waves, denoted as $\Pi_c$ in~\cite{Li}.  Note that $\Pi_c$ is an invariant manifold of the closed-loop (\ref{closednls}) dynamics because $q_0\in~\Pi_c$, therefore, if $q(t=0)$ is a plane wave, then it will remain so.  In what follows, we will assume $q(t=0) \in \Pi_c$.

We ask under what conditions does $q \rightarrow~q_0$?  We will find, as is demonstrated in~\cite{Li,Haller,LiWiggins}, that two attracting solutions can exist on $\Pi_c$.  For $\epsilon_R \gg a^2$ ($\epsilon_I=0$), however, the whole
complex plane attracts to the node.   We will also show that the open-loop dynamics of (\ref{NLS}) are that of a shear flow and therefore a Takens-Bogdanov bifurcation occurs at the target $q_0$ when the control is applied.  As discussed in Section~\ref{takens}, the presence of the Takens-Bogdanov bifurcation means that as $\epsilon_R \downarrow 0$, the target has a small basin of attraction in the purely dissipative case ($\epsilon_I=0$).  Thus, only a small amount of noise ($\sigma_c \approx O(\epsilon_R^2)$, see below) causes loss of control of the system when a purely dissipative control law is applied.  Once control is lost, the system settles onto the other attractor, which is a plane wave of much smaller amplitude.  In Section~\ref{1DHS} we explained that the presence of the conservative term, $\epsilon_I$, will cause the basin of attraction for $q_0$ to increase in size, hence, increasing the controllability of the system.  In what follows we will demonstrate that effect as well.  We will finish our analysis in Section~\ref{concl} with a brief discussion of control of NLS to spatially non-uniform targets.  This is still work in progress and will be treated in a separate paper.

Before we move on, it is worth mentioning that Friedland has shown that the NLS can be autoresonantly excited~\cite{Lazar1,Lazar2} and controlled.  Autoresonance occurs when a nonlinear 
oscillating system phase locks to a small amplitude oscillating drive with a slow frequency chirp.  Autoresonance results in self-consistent control of the amplitude of the system as the drive frequency changes because the driven system changes its state in space and/or time in order to phase lock to the drive. For systems like NLS, where the frequency is 
a function of amplitude, this means that phase locking can be used to manipulate the amplitude \textit{without feedback} and using a small gain (coupling), $\epsilon$.  In~\cite{Lazar1}, Friedland and Shagalov demonstrate that the plane wave state of the NLS can be autoresonantly excited, and that as the amplitude reaches a certain threshold, a spatially modulated form arises and eventually becomes a shape not unlike a soliton.  In~\cite{Lazar2}, Friedland extends his work to standing waves, and more recently  with Shagalov~\cite{Lazar3} has shown how to excite multi-phase solutions of the Korteweg-de Vries (KdV) equation.  Our work complements that done in~\cite{Lazar1,Lazar2,Lazar3}:  their work deals with a drive 
with fixed gain and a frequency chirp, ours has a drive with a fixed `target' (no chirp) and we consider the size of the basin of attraction.

\subsection{Analysis}
Restricting ourselves to plane waves and using $q_0(t)=a\exp(2ia^2t)$ as our control target, (\ref{closednls}) becomes:
\begin{equation} \label{NLSclosed2}
iq_t+2|q|^2q=i(\epsilon_R+i\epsilon_I)(a\exp(2ia^2t)-q).
\end{equation}

We begin our analysis by writing $q(x,t)$ in the form, $q(t)=\rho(t)e^{i\theta(t)}$ (where $\rho$ and $\theta$ are real functions of time) and subsituting it into (\ref{NLS}) to study the open-loop dynamics.  Upon substitution into (\ref{NLS}) we get,
\begin{eqnarray} \label{rhophi}
\dot \rho  &= &  0, \nonumber\\
\dot \theta  &= &  2\rho^2. 
\end{eqnarray}

\begin{figure}[ht]
\begin{center}
\scalebox{0.4}{{\includegraphics{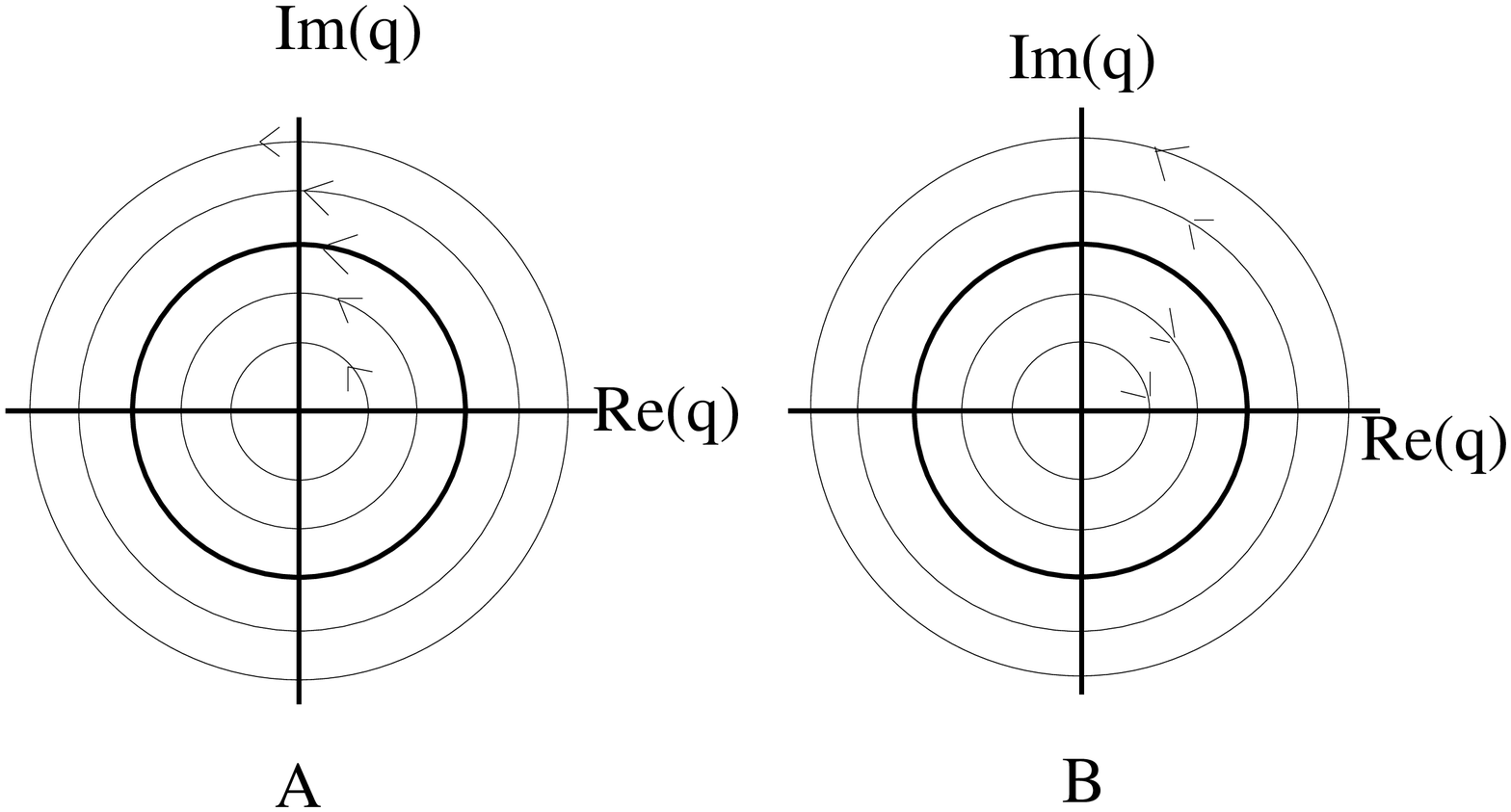}}}
\end{center}
\caption{The dynamics for (\ref{rhophi}) when $\epsilon_R=\epsilon_I=0$ are shown in Figure~\ref{figure:rhophipic}A, note that this is a shear flow.  The bold circular orbit has $|q|=a$.  Figure~\ref{figure:rhophipic}B shows the dynamics of (\ref{rhophi}) transformed into coordinates rotating with the target.  Notice that Figure~\ref{figure:rhophipic}B is a shear flow with a fixed circle of netural stability.}  
\label{figure:rhophipic}
\end{figure}

As shown in Figure~\ref{figure:rhophipic}A, (\ref{rhophi}) describes a shear flow as expected from Section~\ref{1DHS}.  We can also transform into coordinates rotating with the target (the bold circle in Figure~\ref{figure:rhophipic}A) of the closed-loop dynamics by setting $\psi(t)=2a^2t-\theta(t)$ and looking at the $\rho-\psi$ dynamics as shown in Figure~\ref{figure:rhophipic}B where $\dot \psi=2(a^2-\rho^2)$.  Thus relating the dynamics of NLS to (\ref{gen}).

Next, we analyze the closed-loop dynamics by inserting our ansatz, $q(t)=\rho(t)e^{i\theta(t)}$, into (\ref{NLSclosed2}) to get,
\begin{eqnarray} \label{rhophiclosed}
\dot \rho & = & \epsilon_R(a\cos(\psi) - \rho) - \epsilon_Ia\sin(\psi), \nonumber\\
\dot \psi & = & 2(a^2-\rho^2) - \frac{a\epsilon_R}{\rho}\sin(\psi)-\frac{\epsilon_I}{\rho}(a\cos(\psi)-\rho). 
\end{eqnarray}
\begin{figure}[ht]
\begin{center}
\scalebox{0.4}{{\includegraphics{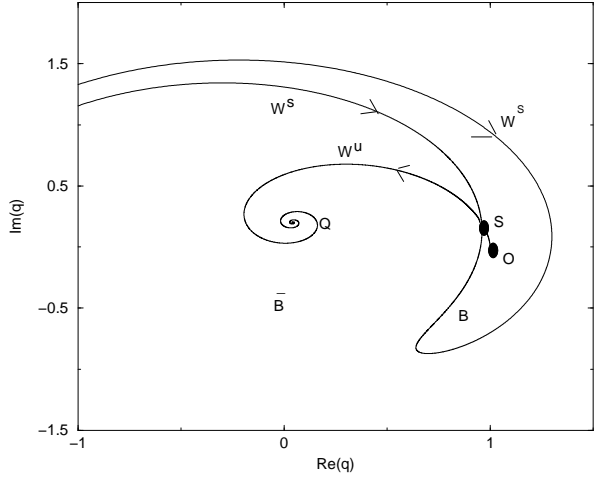}}}
\end{center}
\caption{
The state space diagram for (\ref{rhophiclosed}) with $a=1$, $\epsilon_I=0$, and $\epsilon_R=0.4$.  The variables are $x=Re(q)$ and $y=Im(q)$.  Notice how the stable manifold ($W^s$) of the saddle ($S$) creates the tear drop shaped basin of attraction, $B$, for the target, $O$.  The unstable manifold of the saddle ($S$) is denoted by $W^u$.  The large value of $\epsilon_R$ was chosen to accentuate the features of the basin.   }
\label{figure:xyepr}
\end{figure}

Figure~\ref{figure:xyepr} illustrates the state space of (\ref{rhophiclosed}).  The variables are the real ($x=\rho\cos(\psi)$) and imaginary ($y=\rho\sin(\psi)$) parts of $q$ with $\epsilon_I=0$ and $\epsilon_R=0.4$.  The tear drop shaped basin of attraction, $B$, for the target, $O$, is clearly present and is characteristic of the Takens-Bogdanov bifurcation.  Notice how small $B$ is locally, even though the strength of the dissipation is quite high ($\epsilon_R=0.4$).  Physically, this tells us that only a small region of ``nearby" states are controllable to our target.  Worse yet, there is a small noise threshold for instability.  This can be quantified by measuring the shortest distance, $\sigma_c(\epsilon_R)$, between the basin boundary , $W^s$, and the target, $O$, in the negative $x$-direction.  The point in Figure~\ref{figure:xyepr} denoted $S$ is a saddle point whose location will play a crucial role in computing $\sigma_c$ as was shown in Section~\ref{takens}.  The unstable manifold of the saddle, $S$, is denoted $W^u$. The point, $Q$, is a stable spiral and is associated with a small amplitude plane wave.

\begin{figure}[ht]
\begin{center}
\scalebox{0.4}{{\includegraphics{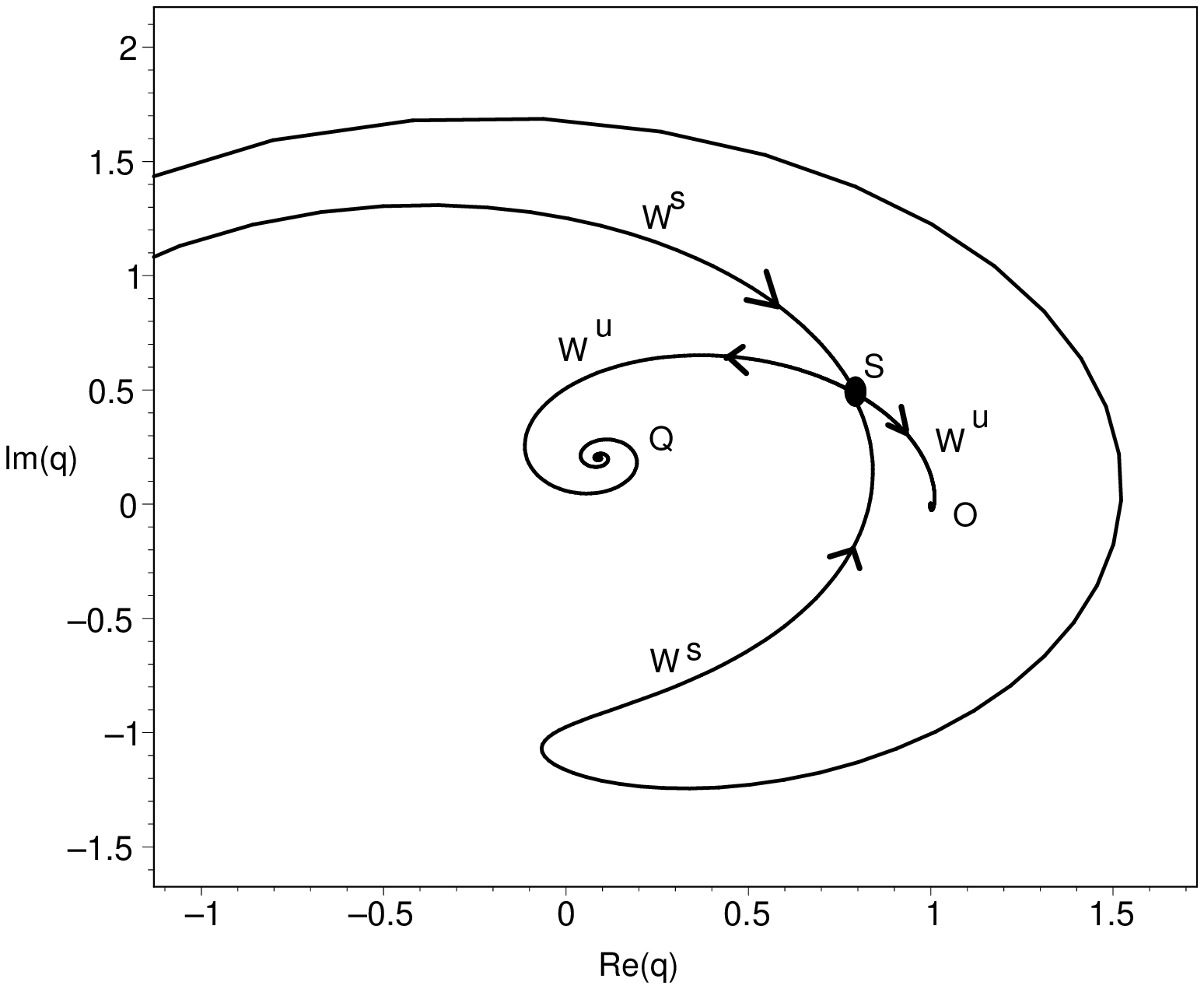}}}
\end{center}
\caption{The state space diagram for (\ref{NLSclosed2}) with $a=1$, $\epsilon_I=-0.1$, and $\epsilon_R=0.4$.  Notice that the tear drop shaped basin of attraction is still present, but is larger than in Figure~\ref{figure:xyepr}.  Even though $\epsilon_I$ is small compared to $\epsilon_R$, its effect is quite noticeable.}
\label{figure:cons}
\end{figure}

In Section~\ref{1DHS} we stated that the presence of the conservative term in the controller ($\epsilon_I$) will enlarge the basin, $B$.  Figure~\ref{figure:cons} illustrates this effect.   Figure~\ref{figure:cons} shows the state space of (\ref{NLSclosed2}) with $\epsilon_R=0.4$ and $\epsilon_I=-0.1$.  Notice how much larger $B$ is in Figure~\ref{figure:cons} as compared to Figure~\ref{figure:xyepr}.  The labels in Figure~\ref{figure:cons} denote the same points and manifolds as in Figure~\ref{figure:xyepr}.  It is interesting to note that the conservative term need not be large in order for its effect on $\sigma_c$ to be noticeable.  

As is implied in Section~\ref{1DHS}, the most useful set of coordinates will be the action-angle coordinates.  The action can be found using $(q^*,q)$ as our dynamical variables, $(p,q)$, and integrating $pdq$ around one cycle (noting that the period, $T=2\pi/\omega=\pi/\rho^2$)~\cite{Arnold,percival,goldstein},
\begin{equation}
I=\frac{1}{2\pi}\oint pdq=\frac{1}{2\pi}\oint p\frac{dq}{dt}dt=\frac{1}{2\pi i}\int_{t=0}^{\pi/\rho^2} q^*2i|q|^2qdt=\rho^2.
\end{equation} 

We can then rewrite $q(t)$ in terms of action-angle coordinates $(I, \phi)$ by $q(t)=\sqrt{I}\exp(i \phi)$ with the target action, $I_0=a^2$.  Using (\ref{NLSham}) we can re-write the Hamiltonian,
\begin{eqnarray}
H(I) & = & |q|^4=I^2, \nonumber \\
\dot{  \phi} & = & 2I.
\end{eqnarray}

In Section~\ref{1DHS}, we transformed our angle coordinate into a new coordinate rotating with the target angle using a generating function, $F_2(I',\phi,t)$.  Recall that this transformation does not change the action coordinate.  Following similar arugments presented in Section~\ref{1DHS} we find,  
\begin{eqnarray}
F_2 & = & (\phi-2I_0t)(I'+I_0), \nonumber \\
 I & = & I' + I_0, \nonumber \\
K(\tilde I) &=&H(I'+I_0) - 2I_0(I_0+I'), \nonumber \\
\Phi & = & \phi - 2I_0 t.
\end{eqnarray} 
This transformation is equivalent to substituting $q(t)=w(t)\exp(2ia^2t)$ into (\ref{NLSclosed2}),
\begin{equation} \label{wnls}
iw_t-2a^2w+2|w|^2w=i(\epsilon_R+i\epsilon_I)(a-w).
\end{equation}
We can write $w(t)$ in terms of action-angle coordinates $w(t)=\sqrt{I'}\exp(i\Phi)$, substitute this into(\ref{wnls}) and expand about the target, $I'=a^2+y$ and $\Phi = 0 + x$ to get,
\begin{equation} \label{linearized}
\left (
\begin{array}{c}
\dot x \\
\dot {y}
\end{array}
\right ) =
\left (
\begin{array}{cc}
-\epsilon_R & 2-\frac{\epsilon_I}{2a^2} \\
2\epsilon_Ia^2 & -\epsilon_R
\end{array}
\right )
\left (
\begin{array}{c}
x \\
y
\end{array}
\right )+
\left (
\begin{array}{c}
\frac{\epsilon_I}{2}x^2 - \frac{3\epsilon_I}{8a^4} {y}^2 + \frac{\epsilon_R}{2a^2} xy \\
{-\epsilon_R a^2}x^2 - \frac{\epsilon_R}{4a^4} {y}^2 + {\epsilon_I} xy
\end{array}
\right ),
\end{equation}
note that $(x,y)$ are now linearization variables and not the variables from Figure~\ref{figure:xyepr}.  Further, note that from Section~\ref{takens}, we know that the $x^2$ term in the $\dot{y}$ equation is the dominant nonlinearity.  After performing a near-identity transformation, the normal form of (\ref{linearized}) is:
\begin{equation} \label{nlsnode}
\left (
\begin{array}{c}
\dot x \\
\dot {y}
\end{array}
\right ) =
\left (
\begin{array}{cc}
-\epsilon_R & 2 \\
2\epsilon_Ia^2 & -\epsilon_R
\end{array}
\right )
\left (
\begin{array}{c}
x \\
y
\end{array}
\right )+
\left (
\begin{array}{c}
0 \\
{-\epsilon_R a^2}x^2 
\end{array}
\right ) + \ldots .
\end{equation}
Note, the near-identity transformation does not effect the leading order (in terms of $\epsilon_R$ and $\epsilon_I$) nonlinear terms.   Further, we notice that in order for an island to open around the target, $\epsilon_I$ must be negative because $\eta=2a^2>0$.  From now on, we insert the negative sign explicitly.  In this case, the bifurcation is exactly as discussed in Section~\ref{takens}.   

\begin{figure}[ht]
\begin{center}
\scalebox{0.4}{{\includegraphics{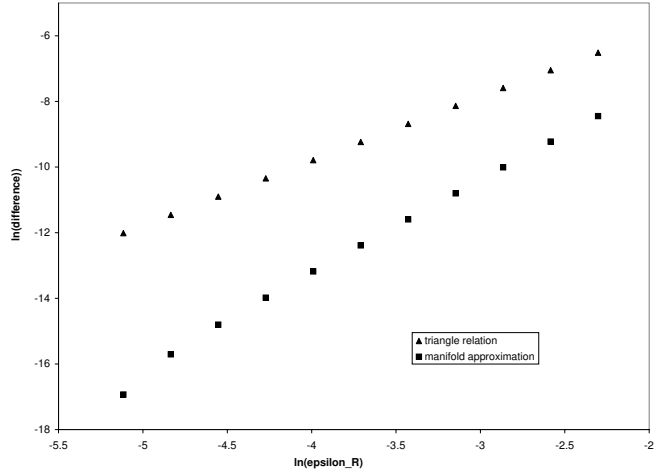}}}
\end{center}
\caption{
Relative error in predicting $\sigma_c$.  The triangles denote the error using the first order Taylor (\textit{i.e.} triangle relation) approximation of $\sigma_c$ and the squares denote the error using the second order Taylor approximation of $\sigma_c$.  Here, relative error means the magnitude of the difference between the numerical result of $\sigma_c$ and the $\sigma_c$ predicted by the two methods of approximation.}
\label{figure:verify}
\end{figure}

Next, we find the position of the saddle:
\begin{equation}
{x}_S = -\frac{{\epsilon_R}^2+4a^2\epsilon_I}{2a^2 \epsilon_R},\mbox{ } y_S=-\frac{{\epsilon_R}^2+4a^2\epsilon_I}{4a^2},
\end{equation}
and we enforce, $\epsilon_I \ll \epsilon_R$, for the reason discussed previously.
Following Section~\ref{takens}, the next step is to find the dynamics of (\ref{nlsnode}) linearized about the saddle with $x={x}_S+u$ and , $y=y_S+v$,
\begin{equation} \label{nlssad}
\left (
\begin{array}{c}
\dot u \\
\dot {v}
\end{array}
\right ) =
\left (
\begin{array}{cc}
-\epsilon_R & 2 \\
{\epsilon_R}^2+2a^2\epsilon_I & -\epsilon_R
\end{array}
\right )
\left (
\begin{array}{c}
u\\
v
\end{array}
\right ).
\end{equation}
Next, we find the angle, $\theta$, between the eigenvectors of (\ref{nlssad}).  Using the formula for $\theta$ from Section~\ref{takens}, we find that:
\begin{equation}
\theta =\frac{\sqrt{2}}{2}\sqrt{2\epsilon_R^2+4a^2\epsilon_I}, 
\end{equation} 
hence, using the triangle relation we find:
\begin{equation} \label{nlssigma}
\sigma_c \approx |x_s\theta| =\frac{\sqrt{2}}{2}\frac{{\epsilon_R}^2+4a^2\epsilon_I}{2a^2 \epsilon_R}\sqrt{2\epsilon_R^2+4a^2\epsilon_I}, \end{equation} 
where $\epsilon_I \ll \epsilon_R$.  The higher order corrections are developed in Appendix~\ref{appendix:error}. We see that the basin boundary in the purely dissipative case is only a distance of $O(\epsilon_R^2)$ away (because for NLS $b_1=-a^2\epsilon_R$).   In \textit{standard} saddle-node bifurcations, $\sigma_c$ scales like $O(\epsilon_R)$.  In Figure~\ref{figure:verify}, we compare the results for finding $\sigma_c$ using the triangle method with the results from an approximation of the stable manifold using higher order terms in the Taylor series expansion (see Appendix~\ref{appendix:error} for more details) for the purely dissipative case.  
Figure~\ref{figure:verify} plots the difference between the numerical value of $\sigma_c$ and the value estimated by each of the two methods.  The numerical value of $\sigma_c$ is found by computing where the stable manifold intersects the $x$-axis of the state space (similar to Figure~\ref{figure:xyepr}).  It is important to note that the equations integrated for Figure~\ref{figure:verify} are not (\ref{linearized}), but rather the equations of motion for the real and imaginary parts of (\ref{NLSclosed2}).  For more details on this, please consult Appendix~\ref{appendix:xyderive}.  Figure~\ref{figure:verify} shows that the error in both methods converges to zero, however, the manifold approximation method does so much more quickly.  Even though the normal form analysis is performed in the limit $\epsilon_R \rightarrow 0$ and, hence, the predictions are only valid in this limit, Figure~\ref{figure:verify} shows the purely dissipative scaling of $\sigma_c \approx \epsilon_R^2$, predicted by the triangle relation, is good for a large range of $\epsilon_R$'s.  
  
\section{Conclusions} \label{concl}
We have shown that when attempting to control an integrable Hamiltonian system to one of its exact solutions a succesful control law will contain both dissipative and conservative terms.  One source of the difficulty for control is the degenerate shear flow structure inherent to these systems.  The shear flow sets up a Takens-Bogdanov bifurcation which leads to the small basin of attraction for the control target.  The presence of the conservative term enlarges the basin and therefore improves the controllability of the system.  Successful control laws for integrable Hamiltonian systems are ones which are more \textit{conservative} than \textit{dissipative}, but dissipative terms must be present in order for the system to settle onto the control target and must be strong enough to allow for a relatively short decay time.

Selection of the NLS to illustrate our results was more than a choice of convenience.  We plan to use the insights gained in controlling NLS to develop control laws for non-integrable systems related to NLS: the previously mentioned Ginzburg-Landau and Dysthe's equation serve as examples of such systems.  These NLS-like systems contain physical phenomenon  which destroys the integrability.  Such control laws may suppress or excite pulse formation in the system, for example.  We have had success in controlling both Ginzburg-Landau and Dysthe's equation to NLS targets, therefore we know that studying integrable Hamiltonian models has relevance to the development of control laws for non-integrable systems.  These targets are spatially non-uniform and are the $\theta$-function solutions from~\cite{GeneNLS}.  This work is still incomplete and will be presented elsewhere.  The high dimensionality of these systems makes detailed analysis difficult and led to our thorough study of the plane wave presented here.    

Future work in this area could include studying the control of integrable Hamiltonian systems with $N$ degrees of freedom ($2N$-dimensional).  The behavior of higher dimensional systems has the potential to be much more complex than the case studied in this paper.  To illustrate this, consider a $2N$-dimenional integrable Hamiltonian system with Hamiltonian, $H(\mathbf{I})$, where $\mathbf{I} \in \mathbb{R}^N$.  As in Section~\ref{1DHS}, we transform to a coordinate system which fixes some target solution, $(\mathbf{I}_0,\bm{\phi_0})$, at the origin.  The Hamiltonian now becomes:
\begin{equation}
K(\mathbf{I}') = \frac{1}{2}\mathbf{I}' \cdot C \cdot \mathbf{I}' + h(\mathbf{I}'),
\end{equation}
where: we have ignored the constant terms, $h$ contains terms of $O(\mathbf{I}'^3)$ and higher, and the $N \times N$ matric, $C$, is the Hessian:
\begin{equation}
C_{jk} \equiv \frac{\partial^2 H}{\partial {I'}_j \partial {I'}_k}|_{\mathbf{I}_0}.
\end{equation}.  
Because $C$ is a real symmetric matrix, we can diagonalize it with a similarity transformation using a rotation matrix, $R$.  This leads to the new system:
\begin{eqnarray}
K'(\mathbf{A}) & = & \frac{1}{2}\mathbf{A} \cdot D \cdot \mathbf{A} + h'(\mathbf{A}), \nonumber \\
D & = & diag(\lambda_1, \lambda_2, \ldots, \lambda_N), \nonumber \\
\dot{{A}}_k &  = & 0, \nonumber \\
\dot \Theta_k & = & \frac{\partial K'(\mathbf{A})}{\partial A_k},
\end{eqnarray}
where $k=1,\ldots,N$, $(\mathbf{A},\bm{\Theta})$ are the new coordinates, and $\lambda_1,\ldots,\lambda_N$ are the eigenvalues of $C$.  Afer re-ordering the basis, the new equations of motion are:
\begin{equation} \label{ndopen}
\left ( \begin{array}{c} \dot \Theta_1 \\ \dot A_1 \\ \dot \Theta_2 \\ \dot I_2 \\ \vdots \\ \dot \Theta_N \\ \dot A_N \end{array} \right ) = 
\left ( \begin{array}{cccccccc}
0 & \lambda_1 & 0 & 0         & 0 & \cdots & 0 & 0 \\
0 &  0        & 0 & 0         & 0 & \cdots & 0 & 0  \\
0 & 0         & 0 & \lambda_2 & 0 & \cdots & 0 & 0 \\
0 & 0		  & 0 & 0		&  0 & \cdots & 0 & 0 \\
\vdots & \vdots & \vdots &  \vdots & \vdots & \ddots & \vdots & \vdots \\
0 & 0        & 0 & 0         & 0  &  \cdots & 0 & \lambda_N \\
0 & 0        & 0 & 0         & 0  &  \cdots & 0 & 0
\end{array} \right )
\left ( \begin{array}{c} \Theta_1 \\  A_1 \\  \Theta_2 \\  I_2 \\ \vdots \\  \Theta_N \\  A_N \end{array} \right ) + 
\left ( \begin{array}{c} g_1(\mathbf{A}) \\ 0 \\ g_2(\mathbf{A}) \\ 0 \\ \vdots \\ g_N(\mathbf{A}) \\ 0 \end{array} \right ),
\end{equation}
where the $g_i$ contains terms of $O(\mathbf{A}^2)$ and higher.  The system of equations (\ref{ndopen}) are the open-loop dynamics for an $2N$-dimensional integrable Hamiltonian system whose origin lies at the control target. The linear term of (\ref{ndopen}) is separated into $N$ non-interacting subspaces whose dynamics are those of a shear flow, as expected.  This potentially leads to Takens-Bogdanov bifurcations of very high dimensionality.  Knowing how the target's basin scales (after control is applied) is crucial for understanding the control of the higher dimensional systems.  The high dimensionality makes an analysis which is similar to that done in Section~\ref{takens} difficult.  Hence this problem is still open to study.  

\begin{acknowledgments}
The authors would like to thank the US Department of Energy's Office of Fusion Energy Sciences for supporting this work.  We would also like to thank Lazar Friedland, Alfred Osborne, Allan Kaufman, Bill Cooke, and the nonlinear studies group at The College of William and Mary for helpful comments and suggestions.  
\end{acknowledgments}

\appendix

\section{The Transformed Control Law} \label{2ndapp}

Consider (\ref{closedloop}):
\begin{equation} \label{cl1}
\dot{ \mathbf{z}} = J\nabla_\mathbf{z}H(\mathbf{z}) + \epsilon\cdot (\mathbf{z}_0(t)-\mathbf{z}),
\end{equation}
where $\epsilon=\epsilon_R1+\epsilon_IJ$.  We transform to a new set of coordinates, $\mathbf{Z}\in \mathbb{R}^{2N}$, via a time dependent transformation on $\mathbf{z}=\mathbf{z}(\mathbf{Z};t)$ such that $\mathbf{Z}(0;t)=\mathbf{z}_0(t)$.  We first find $\dot{\mathbf{Z}}=d\mathbf{Z}/dt$ following an arbitrary orbit $\mathbf{z}(t)=\mathbf{z}(\mathbf{Z}(t);t)$.  Define the Jacobian of the transformation to be:
\begin{equation} \label{jacobian}
M_{ki} \equiv  \frac{\partial z_k}{\partial Z_i} , 
\end{equation}
and note that $M=M(\mathbf{Z};t)$.  Further, we denote the Jacobian following the target orbit as $M_0(t)=M(0;t)$.  We also assume that the transformation is smooth and invertible (\textit{i.e.} $\det(M) \ne 0$ for all $\mathbf{Z}$ and $t$).  The time derivative of $\mathbf{z}$ becomes:
\begin{eqnarray} 
\frac{dz_k}{dt} & = & \frac{\partial z_k}{\partial Z_l}\frac{dZ_l}{dt} + \frac{\partial z_k}{\partial t},  \nonumber \\
\dot{\mathbf{z}} & = & M\dot{\mathbf{Z}} + \frac{\partial \mathbf{z}}{\partial t}, \label{timederiv}
\end{eqnarray}
where summation over repeated indices is implied.
Equation (\ref{timederiv}) holds for \textit{any orbit}.  Note that since $\mathbf{Z}=0$ is a fixed point, we have:
\begin{equation}
\dot{\mathbf{z}}_0 = \frac{\partial \mathbf{z}}{\partial t}|_{(\mathbf{Z}=0;t)}.
\end{equation}
The gradient transforms as:
\begin{equation}
\frac{\partial}{\partial z_k}  =  \frac{\partial Z_m}{\partial z_k}\frac{\partial}{\partial Z_m}.
\end{equation}
Therefore,
\begin{equation}
  J_{kl}\frac{\partial}{\partial z_l}  =  J_{kl} \frac{\partial Z_m}{\partial z_l}\frac{\partial}{\partial Z_m}.
\end{equation}
Noting that $M^{-1}_{ml} = \partial Z_m / \partial z_l$ and substituting the above results into (\ref{cl1}) we get:
\begin{equation} \label{newlaw}
\dot{\mathbf{Z}} = M^{-1}J\tilde{M}^{-1}\nabla_\mathbf{Z}H(\mathbf{z}(\mathbf{Z};t)) - M^{-1}\frac{\partial \mathbf{z}(\mathbf{Z};t)}{\partial t} + M^{-1} \cdot \epsilon \cdot (\mathbf{z}_0(t) - \mathbf{z}(\mathbf{Z};t)),
\end{equation}
where $\tilde M$ denotes the transpose of $M$.

The equation (\ref{newlaw}) is true for a \textit{general} coordinate transformation.   In particular, we are interested in canonical transformations.  In canonical transformations,  $MJ\tilde{M} = J$ for all $\mathbf{Z}$ and we say that $M$ is \textit{symplectic}.  Because the set of symplectic matrices form a group, the symplectic condition also holds for $M^{-1}$ for any $M$ in the symplectic group.   We further note that from the theory of canonical transformations~\cite{goldstein}, the $M^{-1} \partial \mathbf{z} / \partial t$ term can be re-cast in Hamiltonian form and the open-loop dynamics written as $\dot{\mathbf{Z}}=J\nabla_\mathbf{Z}K(\mathbf{Z},t)$ for some new time dependent Hamiltonian, $K$.  What interests us most is the form of the control terms (with control gain $\epsilon$) under the transformation.  We find this by
Taylor expanding $\mathbf{z}(\mathbf{Z};t)$ about the target, $\mathbf{z}_0$,
\begin{eqnarray}
\mathbf{z}(\mathbf{Z};t) &=& \mathbf{z}_0(t) + M_0(t) \cdot \mathbf{Z} + O(\mathbf{Z}^2), \nonumber \\
M(\mathbf{Z};t) &=& M_0(t) + O(\mathbf{Z}), \nonumber \\
M^{-1} & = & {M_0}^{-1}(t) + O(\mathbf{Z}).
\end{eqnarray}
The control term then becomes:
\begin{eqnarray}
M^{-1} \cdot \epsilon \cdot (\mathbf{z}_0(t) - \mathbf{z}(\mathbf{Z};t)) & = -& M_0^{-1} \cdot \epsilon \cdot M_0 \cdot \mathbf{Z} + O(\mathbf{Z}^2), \nonumber \\
M_0^{-1}(\epsilon_R 1 + \epsilon_I J) M_0 & = & \epsilon_R 1 + \epsilon_I M_0^{-1}JM_0, \nonumber \\
&=& \epsilon_R 1 + \epsilon_I J \tilde{M_0}M_0,
\end{eqnarray}
where we have used the symplectic condition to write ${M_0}^{-1}J=J\tilde{M}_0$.

If we define $S_0(t) =  \tilde{M_0}M_0$, (\ref{newlaw}) becomes:
\begin{equation} \label{newclosed}
\dot{\mathbf{Z}} = J\nabla_\mathbf{Z}K(\mathbf{Z};t)  -\epsilon_R 1\cdot \mathbf{Z} - \epsilon_I J S_0(t) \cdot \mathbf{Z} + O(\mathbf{Z}^2),
\end{equation}
as quoted in the text.

\section{Derivation of the Island Width for Conservative Control} \label{1stapp}
The typical scaling of the island width with purely conservative control can be most easily demonstrated by example.  Consider a nonlinear one-dimensional Hamiltonian system with Hamiltonian:
\begin{equation}
H = \frac{1}{2}I^2 + h(I),
\end{equation}
where $h(I)$ contains terms of $O(I^3)$ and higher.  In this system, the target oribit is $(I=0,\phi=\omega_0t)$.  We add a general conservative perturbation, which is periodic and resonant with the target:
\begin{equation} \label{kam}
K'(I,\phi) = \frac{1}{2}I^2-a\epsilon_I \cos(\phi) + h(I),
\end{equation}
the appearance of the minus sign in front of the cosine term ensures that the island opens at the target, as explained in the text.  It is easy to verify that the fixed points for the system with Hamiltonian, $K'$, are at ($I=0,\phi=n\pi$) for $n=0,1,2,\ldots$.  Figure~\ref{figure:island} illustrates the phase space for our controlled system (\ref{kam}), while Figure~\ref{figure:AAdynamics}B shows the open-loop dynamics for this system. The value of $K'$ along the separatrix is given by its value at the saddle point, $K'(0,-\pi)=a\epsilon_I$.  Therefore, along the separatix
\begin{equation} \label{Imax}
\frac{1}{2}I^2  - a\epsilon_I \cos(\phi) = a\epsilon_I.
\end{equation}
Solving (\ref{Imax}) for $I_S(\phi)$, the action $I_S(\phi)$ is at a maximum when $\phi=0$, therefore it is easy to solve for the width of the island, $\Delta\approx O(\sqrt{a\epsilon_I})$. 

\section{Higher Order Estimates of $\sigma_c$}\label{appendix:error}

In this Appendix, we will examine an approximate equation for the stable manifold of the saddle in order to obtain an error estimate for (\ref{newscale1}) and for (\ref{nlssigmaxy}).  This will be done using a Taylor approximation for the stable manifold of the saddle.  In Section~\ref{takens}, we mentioned that the triangle relation is a first order Taylor series approximation for the stable manifold.  The first order Taylor approximation is, of course, a striaght line.  As can be seen from Figure~\ref{figure:xyepr}, the stable manifold has a curvature to it which requires higher order terms in the Taylor approximation for a more accurate measurement of $\sigma_c$.  We will see that while the triangle relation doesn't not take into account this curvature, it still yields the correct leading order dependence in $\epsilon_R$ for $\sigma_c$.  For simplicity, we will consider only the purely dissipative case ($\epsilon_I=0$).

\subsection{Approximation of the Stable Manifold of (\ref{conservmodel})}
Recall that our generic two-dimensional dynamics about the target are:
\begin{equation} 
\left ( \begin{array}{c} \dot \Phi \\ \dot I'  \end{array} \right ) =
\left ( \begin{array}{cc}
-\epsilon_R & \lambda \\
0 & -\epsilon_R
\end{array} \right )
\left ( \begin{array}{c} \Phi \\ I' \end{array} \right ) +
\left ( \begin{array}{c} 0 \\ b_1 \Phi^2 + b_2 \Phi I' \end{array} \right ).
\end{equation}
Finding an approximate equation for the stable manifold of the saddle is the same as finding an approximate equation for $I'(\Phi)$ near the saddle point.   Hence, we are interested in solving the equation,
\begin{equation} \label{iofphi}
F(\Phi;I') \equiv \frac{dI'}{d\Phi} =  \frac{-\epsilon_RI' + b_1\Phi^2 + b_2I'\Phi}{-\epsilon_R\Phi + \lambda I'},
\end{equation}
for $I'(\Phi)$.  We can think of $F(\Phi;I')$ as a function of the angle, $\Phi$, parameterized by the action $I'$.  The stable manifold will be a curve, $I'(\Phi)$, which passes through the saddle point in the direction of the saddle's stable eigenvector.

Next, we approximate $I'(\Phi)$ by a quadratic polynomial (Taylor series to second order):
\begin{equation} \label{taylor} 
I'(\Phi) \approx \alpha_1 + \alpha_2 \Phi + \alpha_3 \Phi^2.
\end{equation}
We insert the approximation into the left hand side of (\ref{iofphi}) and Taylor expand the right hand side of (\ref{iofphi}) about the saddle:
\begin{equation}
\alpha_2 + 2\alpha_3 = -\frac{(2\lambda b_1+\epsilon_R b_2)\epsilon_R}{(\lambda b_1+\epsilon_R b_2)\lambda} - \frac{b_1}{2\epsilon_R}\Phi.
\end{equation}
By equating terms of similar powers of $\Phi$ gives:
\begin{eqnarray}
\alpha_2 & = & -\frac{(2\lambda b_1+\epsilon_R b_2)\epsilon_R}{(\lambda b_1+\epsilon_R b_2)\lambda}, \nonumber \\
\alpha_3 & = & -\frac{b_1}{2\epsilon_R}. \nonumber \\
\end{eqnarray}
The term $\alpha_1$ is found by observing that:
\begin{equation}
I'_S(\Phi_S) = \alpha_1 + \alpha_2 \Phi_S + \alpha_3 \Phi_S^2, 
\end{equation}
this gives:
\begin{equation}
\alpha_1 = \frac{2\epsilon_R^3}{(b_1\lambda+\epsilon_Rb_2)\lambda}\frac{7\lambda b_1 + 4b_2\epsilon_R}{4(b_1\lambda+\epsilon_R b_2}.
\end{equation}
We can approximate $\sigma_c$ by finding where the stable manifold intersects the $\Phi$-axis (see Figure~\ref{figure:manifold} below, where locally the $x$ coordinate is analagous to $I'$ and $y$ is locally analagous to $\Phi$).  Hence:
\begin{equation} \label{taylorsigma}
\sigma_c \approx \alpha_1 = \frac{2\epsilon_R^3}{(b_1\lambda+\epsilon_Rb_2)\lambda}\frac{7\lambda b_1 + 4b_2\epsilon_R}{4(b_1\lambda+\epsilon_R b_2}.
\end{equation}
We see that (\ref{taylorsigma}) is written in the form:  $\sigma_c \approx \sigma^{(0)}_cg(\epsilon_R)$, where $\sigma^{(0)}_c$ is the noise threshold given by the triangle relation.  Immediately, we see that the triangle relation gives the correct parameter dependence (in $\epsilon_R$).  The multiplicative constant not present in the triangle relation results is due to the curvature of the stable manifold and does not change the leading order parameter dependence.  For a clarification of this, see Figure~\ref{figure:manifold} below.  We can systematically improve this estimate by including more terms in the Taylor expansion (\ref{taylor}).  Next, we will compute $\sigma_c$ for the NLS problem.

\subsection{Manifold Approximation for NLS}
\begin{figure}[ht]
\begin{center}
\scalebox{0.4}{{\includegraphics{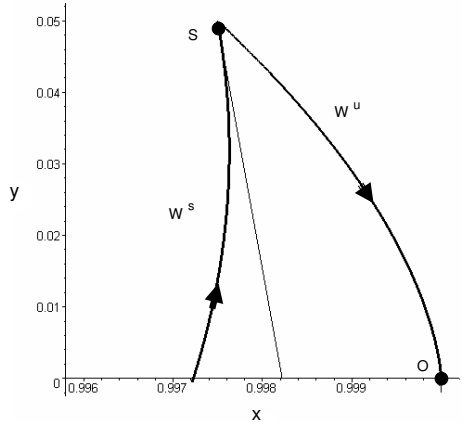}}}
\end{center}
\caption{
The state space of (\ref{xdota}),(\ref{ydota}) near the target. The straight line is the approximation of the stable manifold used by the triangle relation.}
\label{figure:manifold}
\end{figure}

We will use the equations developed in Appendix~\ref{appendix:xyderive} to derive an approximation of the saddle's stable manifold.  We begin with the equations:
\begin{eqnarray}
\dot x & = & \epsilon_R(a-x)  + 2y(x^2+y^2-a^2),   \label{xdota} \\ 
\dot y & = & -\epsilon_Ry  - 2x(x^2+y^2-a^2), \label{ydota}
\end{eqnarray}
where (\ref{xdota}) and (\ref{ydota}) describes the time evolution of the real ($x$) and imaginary ($y$) part of $q$ in (\ref{NLSclosed2}).  The location of the saddle in this coordinate system is:
\begin{equation}
\left ( y_S = \frac{\epsilon_R}{2a}, \mbox{ } x_S = \frac{a^2+\sqrt{a^4-\epsilon_R^2}}{2a} \right ).
\end{equation}
We are interested in finding a curve $x(y)$ which approximates the stable manifold.  As in the eariler case, we study the ratio:
\begin{equation} \label{fxy}
F(y;x) \equiv \frac{dx}{dy} = \frac{\epsilon_R(a-x)+ 2y(x^2+y^2-a^2)}{-\epsilon_Ry  - 2x(x^2+y^2-a^2)}.
\end{equation}
We will approximate the stable manifold with a quadratic polynomial,
\begin{equation} \label{quad}
x(y) \approx \alpha_1 + \alpha_2y + \alpha_3y^2.
\end{equation}
The coefficients $c_2$ and $c_3$ can be using the method outlined above.  We can see from Figure~\ref{figure:xyepr} that $\sigma_c$ lies approximately along the $x$-axis.  Therefore, the most important term in the approximation will be $c_1$.  We find $c_1$ by evaluating (\ref{quad}) at the saddle point:
\begin{equation}
 c_1 \approx a - \frac{5}{18a^3}\epsilon_R^2 +O(\epsilon_R^4) + \ldots.
\end{equation}
Because we approximate $\sigma_c$ as lying along the $x$-axis, $\sigma_c \approx a - c_1$:
\begin{equation} \label{nlssigmafix}
\sigma_c \approx \frac{5}{18a^3}\epsilon_R^2 + O(\epsilon_R^4) + \ldots.
\end{equation}
We see that just like in the generic case, we have a multiplicative correction to the leading order term due to the curvature of the stable manifold (see Figure~\ref{figure:manifold}).  
Figure~\ref{figure:manifold}, shows the state space of (\ref{xdota}),(\ref{ydota}) near the target.  The curve $W^s$ is the stable manifold of the saddle, $S$, and $W^u$ is the unstable manifold which connects the saddle to the target, $O$.  The straight line is the approximation of the stable manifold used by the triangle relation, a first order Taylor approximation of $W^s$.  The second order Taylor approximation of $W^s$ is indistinguisable from $W^s$ at this scale.    We see immediately that (\ref{nlssigmaxy}) gives the correct scaling (as shown in Figure~\ref{figure:verify}) for $\sigma_c$, but (\ref{nlssigma}) is off by a constant.  This can be seen by noting the scale of the $x$-axis in Figure~\ref{figure:manifold}.  In fact, (\ref{nlssigmaxy}) is slightly less than the true value of $\sigma_c$ (which is where $W^s$ intersects the $x$-axis) mainly due to the curvature of $W^s$.  While this plot is generated with a specific value of $\epsilon_R=0.1$, the shape of the manifolds is characteristic of this unfolding of the Takens-Bogdanov bifurcation.

\section{Finding $\sigma_c$ using the Real and Imaginary Parts of (\ref{NLSclosed2})} \label{appendix:xyderive}

In this section, we will present the equations and the analysis used to develop the numerical results of Figure~\ref{figure:verify}.  In Section~\ref{NLSsec}, we chose to use action-angle coordinates because it mirrored the analysis done in Section~\ref{takens}.  However, the more familiar coordinates of $Re(q)$ and $Im(q)$ provide more insight into the physics of the problem.  The $\sigma_c$ used in Figure~\ref{figure:verify} is the one computed below.  We will see that both (\ref{nlssigma}) and (\ref{nlssigmaxy}) have the same scaling in $\epsilon_R$.

We begin by subsituting $q(t)=\rho(t)e^{i(2a^2t-\psi(t))}$ into (\ref{NLSclosed2}):
\begin{eqnarray}
\dot \rho & = & \epsilon_R(a\cos(\psi)-\rho)-\epsilon_Ia\sin(\psi), \\
\rho \dot \psi & = & 2\rho(a^2-\rho^2) -\epsilon_I(a\cos(\psi)-\rho)-\epsilon_Ra\sin(\psi). 
\end{eqnarray} 
These equations now describe the system in a frame rotating with the target solution.  However, note that the target is not at the origin, it is at $(\rho = a , \psi = 0)$.  Next, we transform to cartesian coordinates:  $x=\rho \cos(\psi)$, $y=\rho\sin(\psi)$:
\begin{eqnarray}
\dot x & = & \epsilon_R(a-x) -\epsilon_Iy + 2y(x^2+y^2-a^2),   \label{xdot} \\ 
\dot y & = & -\epsilon_Ry + \epsilon_I(a-x) - 2x(x^2+y^2-a^2). \label{ydot}
\end{eqnarray}
The equations (\ref{xdot}) and (\ref{ydot}) are the equations from which we will derive $\sigma_c$.  The target is located at $(x=a,y=0)$.  For the purposes of Figure~\ref{figure:verify}, we will be only interested in the purely dissipative case.  Hence, we will set $\epsilon_I=0$.  Next, we will compute the normal form about the target to second order (including the $\epsilon_R$ terms) with $y=\beta_1$ and $x=a+\beta_2$:
\begin{equation} \label{xytarget}
\left ( \begin{array}{c} \dot \beta_1 \\ \dot \beta_2 \end{array} \right ) = 
\left(  \begin{array}{cc}
-\epsilon_R & 4a^2 \\
0 & -\epsilon_R
\end{array} \right )
\left ( \begin{array}{c}  \beta_1 \\  \beta_2 \end{array} \right ) + 
\left ( \begin{array}{c}  -2a\beta_1^2 \\ 4a \beta_1 \beta_2 \end{array} \right ) + \ldots.
\end{equation}
We can solve for the location of the saddle using (\ref{xytarget}):
\begin{equation}
\left ( \beta_{1S} = \frac{\epsilon_R}{4a}, \mbox{ } \beta_{2S} = \frac{3\epsilon_R^2}{32a^3} \right ).
\end{equation}
Next, we linearize (\ref{xytarget}) about the saddle, $\beta_1 = \beta_{1S} + \alpha_1$, $\beta_2 = \beta_{2S} + \alpha_2$:
\begin{equation} \label{xysaddledyn}
\left ( \begin{array}{c} \dot \alpha_1 \\ \dot \alpha_2 \end{array} \right ) = 
\left ( \begin{array}{cc}
-2\epsilon_R & 4a^2 \\
\frac{3\epsilon_R^2}{8a^2} & 0 
\end{array} \right )
\left ( \begin{array}{c}  \alpha_1 \\  \alpha_2 \end{array} \right ) + \ldots.
\end{equation}
The angle, $\theta$, between the eigevectors of the linear part of (\ref{xysaddledyn}) can be shown to be:
\begin{equation}
\theta = \frac{\epsilon_R\sqrt{3}}{4a^2}.
\end{equation}
Finally, using the triangle relation, $\sigma_c = \beta_{1S}\theta$, we find:
\begin{equation} \label{nlssigmaxy}
\sigma_c \approx \frac{\epsilon_R^2\sqrt{3}}{16a^3} + \ldots,
\end{equation}
where the error terms have been computed in Appendix~\ref{appendix:error}.

\end{document}